\documentclass[amsmath,amssymb,aps,pra,reprint,superscriptaddress,footnoteinbib,twocolumn,longbibliography]{revtex4-2}
\usepackage{amsmath,amssymb}
\usepackage[english]{babel}
\usepackage{bbold}
\usepackage{upgreek}
\usepackage{dsfont}
\usepackage{graphicx}
\usepackage{comment}
\usepackage{braket}
\usepackage{colortbl}
\usepackage{blindtext}
\usepackage{xr-hyper}
\usepackage[colorlinks]{hyperref}
\hypersetup{colorlinks,linkcolor=blue,citecolor=blue,urlcolor=black,final}
\usepackage{bm}
\usepackage[utf8]{inputenc}
\usepackage{xcolor}
\allowdisplaybreaks
\definecolor{cayenne}{rgb}{0.6, 0, 0}
\usepackage{geometry}
\usepackage{natbib}
\usepackage{xr}

\begin{document}

\title{Gyroscopically stabilized quantum spin rotors}

\author{Vanessa Wachter}
\affiliation{Institute for Complex Quantum Systems and Center for Integrated Quantum Science and Technology, Ulm University, Albert-Einstein-Allee 11, 89069 Ulm, Germany}
\affiliation{Institute for Theoretical Solid State Physics, RWTH Aachen University, 52074 Aachen, Germany}

\author{Silvia Viola Kusminskiy}
\affiliation{Institute for Theoretical Solid State Physics, RWTH Aachen University, 52074 Aachen, Germany}
\affiliation{Max Planck Institute for the Science of Light, Staudtstra\ss{}e 2, 91058 Erlangen, Germany}

\author{Gabriel H\'etet}
\affiliation{Laboratoire de Physique de l'\'Ecole Normale Sup\'erieure, ENS, PSL, CNRS,
Sorbonne Université, Université de Paris, 24 rue Lhomond, 75005 Paris, France}

\author{Benjamin A. Stickler}
\affiliation{Institute for Complex Quantum Systems and Center for Integrated Quantum Science and Technology, Ulm University, Albert-Einstein-Allee 11, 89069 Ulm, Germany}

\begin{abstract}
Recent experiments demonstrate all-electric spinning of levitated nanodiamonds with embedded nitrogen-vacancy spins. Here, we argue that such gyroscopically stabilized spin rotors offer a promising platform for probing and exploiting quantum spin-rotation coupling of particles hosting a single spin degree of freedom. Specifically, we derive the effective Hamiltonian describing how an embedded spin affects the rotation of rapidly revolving quantum rotors due to the Einstein-de Haas and Barnett effects, which we use to devise experimental protocols for observing this coupling in state-of-the-art experiments. This will open the door for future exploitations of quantum spin rotors for superposition experiments with massive objects.
\end{abstract}
\maketitle

The free rotation of a magnetized body can differ significantly from that of the same object if it were nonmagnetic. This, perhaps surprising, discrepancy is due to the fact that magnetic moments carry a spin angular momentum that contributes to the total angular momentum of the magnetized body, as quantified by the material-specific gyromagnetic ratio. Remarkably, it is the {\it total} angular momentum, i.e. the sum of mechanical and spin angular momentum, that enters the equations of motion of the body \cite{vanvleck1951}. This gives rise to the so-called spin-rotation coupling \cite{bunker2006molecular} between rigid-body rotations and magnetic degrees of freedom, as famously witnessed in the Einstein-de Haas and Barnett effects \cite{einstein1915experimental,barnett1915magnetization,frenkel1979history,ganzhorn2016quantum,dornes2019ultrafast}. The impact of spin-rotation coupling typically decreases with particle size, while it is still relevant for microscale and nanoscale magnetic objects \cite{chudnovsky1994conservation,jackson2016precessing,rusconi2017quantum,ma2021torque,rusconi2022spin,kustura2022stability}.

Nanoscale particles of various shapes and with sizes ranging from hundreds of nanometers to tens of microns can be stably suspended in ultrahigh vacuum using optical, electrical, or magnetic traps \cite{gonzalez2021levitodynamics}. Monitoring and controlling their linear and rotational motion with laser fields enables precision sensing of forces and torques as required for fundamental tests of physics \cite{moore2021searching,stickler2021quantum}. Recent demonstrations of cooling their center-of-mass oscillations \cite{delic2020cooling,tebbenjohanns2021quantum,magrini2021real,kamba2021recoil} and their angular librations \cite{pontin2023simultaneous,dania2024high,troyer2025quantum} in an optical tweezer into the quantum regime constitute the first steps toward quantum applications of levitated particles. The demonstrated degree of control and readout renders these systems ideal for performing quantum experiments with objects of unrivaled masses \cite{bateman2014near,llordes2024macroscopic,stickler2018probing,wan2016free,wan2016tolerance,romero2010toward}.

\begin{figure}[b!]
\includegraphics[width=\columnwidth]{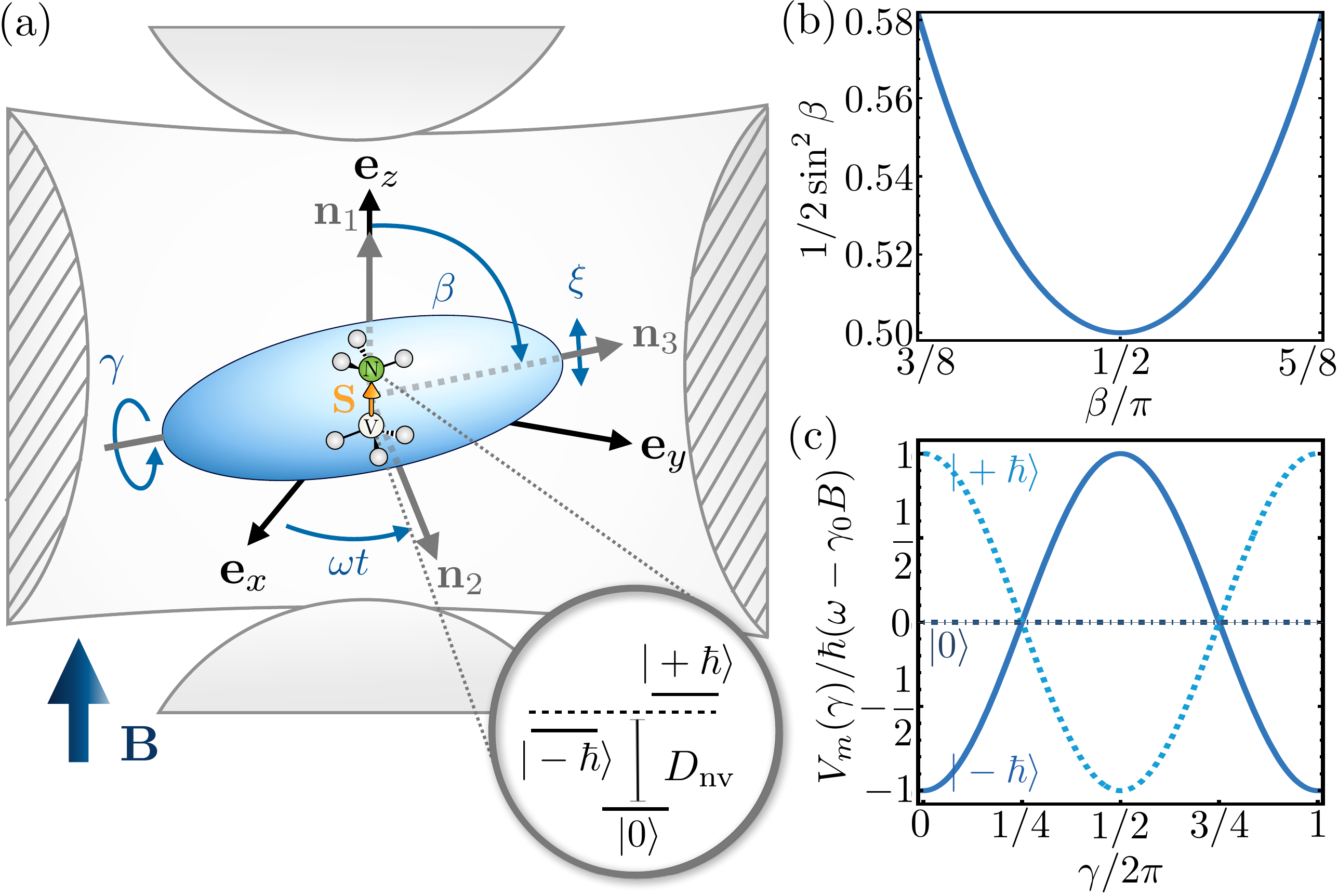}
\caption{(a) Charged spheroidal nanodiamond with an embedded nitrogen-vacancy (NV) center is levitated in a Paul trap and revolves rapidly  with frequency $\omega$ in the symmetry plane of the trap. The  major principal axis $\mathbf{n}_1$ of the body is chosen to be aligned with the NV quantization axis so that the NV spin states $|S_1=\pm\hbar\rangle$ are split by the rotation-induced Barnett field as well as by an applied magnetic field $\mathbf{B}$. The particle rotation induces (b) an effective gyroscopic potential $I\omega^2/2\sin^2 \beta$ stabilizing the out-of-plane rotation angle $\beta$, up to small librations $\xi$, while the external magnetic field  together with spin-rotation coupling induce (c) a spin state-dependent potential $V_{m}(\gamma)$ for $\gamma$ rotations around the body-fixed symmetry axis depending on the spin states $|S_1 =  \hbar m\rangle$ with $m \in\{-1,0,1\}$.}
\label{fig:system}
\end{figure}

Experiments with levitated magnetic objects aim at harnessing their magnetic properties to gain better control over the internal and external degrees of freedom of the nanoparticle  \cite{perdriat2021spin}. Magnetized particles have been trapped in electric \cite{delord2017strong,jin2024quantum}, magnetic \cite{Gutierrez2023,Hofer2023,fuwa2023ferromagnetic}, optical \cite{seberson2019optical,chakraborty2024optomagnetic}, and superconducting traps \cite{gieseler2020single,ahrens2025,jose2025cryogenic}, where their center-of-mass and rotational motion can be monitored and manipulated. Various schemes have been devised to control their mechanical motion via driving the internal magnetization field \cite{gonzalez2020quantumAcoust,wachter2021optical,kani2022magnonic,streltsov2021ground, steiner2024pentacene, ahrens2025observation} or individual magnetic defect centers, such as nitrogen-vacancy (NV) centers \cite{delord2018ramsey,delord2020spin,perdriat2021spin,geiselmann2013three,conangla2018motion,feldman2025trapping,muretova2025parametric,li2024preparing,espinos2024enhancing}. These schemes promise to enable sensing of magnetic fields and accelerations \cite{kumar2017magnetometry,jackson2016precessing,ahrens2025,ni2025microscopic,kalia2024ultralight} as well as the creation of massive superpositions \cite{yin2013large,wan2016free,pedernales2020motional,japha2023quantum,zhou2025gyroscopic,rizaldy2025rotational}, and perhaps even probing the quantumness of gravity \cite{bose2017spin}.

In this Letter, we show how the spin-rotation coupling between a levitated nanoparticle and a single embedded spin degree of freedom can be observed and exploited for quantum experiments. The proposed scheme is based on the recent experimental demonstration of rapid spinning of electrically levitated nanodiamonds with embedded spin degrees of freedom \cite{perdriat2024rotational,jin2024quantum}. This spinning boosts the spin-rotation coupling while it also gyroscopically stabilizes the nanoparticle orientation in the corotating frame (see Fig.~\ref{fig:system}), so a combination of magnetostatic fields and microwave pulses then enables controlling the collective motion of hundreds of millions of atoms via a single spin.

Specifically, we consider a symmetric nanodiamond with a single embedded NV spin levitated in the presence of an external static magnetic field $\mathbf{B}=B\mathbf{e}_z$, as schematically depicted in Fig.~\ref{fig:system}. The nanodiamond is approximated as an ellipsoidal rigid rotor with semiaxis lengths $l_1 =l_2<l_3$ and moments of inertia $I_1 = I_2 = I$ with $I_3 < I$. The nanoparticle orientation $\Omega = (\alpha,\beta,\gamma)$, parametrized by Euler angles in the $z$-$y'$-$z''$ convention, specifies the tensor ${\rm R}(\Omega)$ which rotates from the lab-frame axis ${\bf e}_j$, $j = x,y,z$, to the rotor principal axes $\mathbf{n}_{\mu}(\Omega) = {\rm R}(\Omega) {\bf e}_j$, $\mu = 1,2,3$. For instance, ${\bf n}_1(\Omega) = {\rm R}(\Omega){\bf e}_x$, etc.; an explicit expression for ${\bf n}_\mu(\Omega)$ is given in App.~\ref{app:body-frame}. The inertia tensor depends on the orientation of the particle through $\mathrm{I}(\Omega)=\sum_\mu I_\mu\mathbf{n}_\mu(\Omega) \otimes\mathbf{n}_\mu(\Omega)$. A single NV center with spin-one angular momentum vector operator $\mathbf{S}$ is embedded in the nanodiamond, effectively introducing an internal spin triplet and adding the magnetic dipole moment $\gamma_0 {\bf S}$ to the particle (with gyromagnetic ratio $\gamma_0>0$). The spin angular momentum adds to the mechanical angular momentum to yield the total angular momentum ${\bf J}$ of the system. For the spin quantization axis, we choose $\mathbf{n}_1(\Omega)$, effectively breaking the symmetry of the particle shape; the NV zero-field splitting is $D_{\rm nv} \simeq 2\pi \times 2.87$\,GHz. Possible imperfections, such as weak deviations from axial symmetry or small misalignment of the NV axis relative to the principal inertia axis, are analyzed in App.~\ref{app:imperfections}.

Gyroscopic stabilization is achieved by trapping a charged particle in an azimuthally symmetric quadrupole trap and spinning it with high rotation frequency $\omega$ around the lab-frame ${\bf e}_z$ axis, as recently realized in two experiments \cite{perdriat2024rotational,jin2024quantum}. For homogeneously charged particles the electric quadrupole tensor aligns with the inertia tensor, with quadrupole moments $Q_1 = Q_2 = Q$ and $Q_3 > Q$. For vanishing electric dipole moment, the center-of-mass and rotational motion of the particle decouple so that the resulting rotational macromotion potential is determined by the ac voltage $U_{\rm ac}$, the electrode separation $d_0$, and the ac frequency $\omega_{\rm ac}$ \cite{martinetz2021electric}. The spin-rotor Hamiltonian is
\begin{align}\label{eq:hamiltonian}
    H=& \dfrac{(\mathbf{J}-\mathbf{S})^{2}}{2I}+\left(\dfrac{1}{2I_3}-\dfrac{1}{2I}\right)( J_{3}-S_{3})^{2}+\frac{D_\mathrm{nv}}{\hbar} S^2_1\nonumber\\
&-\gamma_0 B \left( S_1\sin\beta\cos\gamma-S_2\sin\beta\sin\gamma-S_3\cos\beta \right)\nonumber\\
    & +\dfrac{U_\mathrm{ac}^2(Q-Q_{3})^2}{16 I\omega_\mathrm{ac}^2d_0^4}\sin^2\beta \cos^2\beta.
\end{align}
Here, $J_\mu = {\bf J} \cdot {\bf n}_\mu (\Omega)$ denotes the body-frame component of the total angular momentum vector and likewise for ${\bf S}$. The confining quadrupole potential depends only on the polar angle $\beta$, implying that the lab-frame angular momentum component $p_\alpha={\bf J}\cdot {\bf e}_z$ is conserved. It is thus convenient to transform the spin-rotor state $|\psi\rangle$ to a corotating frame of boosted angular momentum via
\begin{equation}\label{eq:trafo_rotFrame}
|\psi\rangle =\exp\left(i\frac{I\omega}{\hbar}\alpha - i \omega t\frac{p_\alpha}{\hbar}  \right )|\phi\rangle.
\end{equation}
This adds the gyroscopic potential $I \omega^2/ 2 \sin^2 \beta$ [see Fig.~\ref{fig:system}(b)] to the quadrupole potential in Eq.~\eqref{eq:hamiltonian}, effectively confining the particle to the rotation plane $\beta \simeq \pi/2$ with harmonic frequency
\begin{equation}
    \omega_\xi = \sqrt{\omega^2 + \dfrac{U_\mathrm{ac}^2(Q-Q_{3})^2}{8 I^2\omega_\mathrm{ac}^2d_0^4}}.
\end{equation}
For large rotation rates $\omega_\xi \simeq \omega$ (see App.~\ref{app:gyropotential}), librations out of the rotation plane can be first linearized in $\xi = \pi/2 - \beta$ and then adiabatically eliminated. Specifically, linearization yields
\begin{align}\label{eq:gyroham}
H_{\rm rot} = & H_\xi + \frac{p^2_\gamma}{2I_3} +\frac{D_{\rm nv}}{\hbar}S_1^2 +(\gamma_0B S_3 - \omega p_\gamma)  \xi \nonumber \\
& + \left ( \omega - \gamma_0B \right )(S_1 \cos\gamma - S_2\sin \gamma ) ,
\end{align}
with $p_\gamma = J_3$, $H_\xi = p_\xi^2/2I + I \omega_\xi^2 \xi^2/2$, and nondominant contributions to spin-rotation coupling as well as dispersion in the gyroscopic motion being neglected (see App.~\ref{app:gyropotential}). The fast librations out of the rotation plane can now be eliminated by neglecting transitions between eigenstates $|n\rangle$ of $H_\xi$ in a Born-Oppenheimer approximation (see App.~\ref{app:adiabaticapproximation}). This is achieved by expanding
\begin{equation}\label{eq:stateexp}
 |\phi\rangle=  \sum_{n = 0}^\infty \exp \left[-\frac{ip_\xi}{\hbar} \frac{\omega p_\gamma - \gamma_0 B S_3}{I \omega^2}\right] e^{-itH_\xi/\hbar}|n,\chi_n\rangle, 
\end{equation}
so that the adiabatic spin-rotor state $|\chi_n\rangle$ evolves according to the Hamiltonian
\begin{figure}[t!]
\includegraphics[width=\columnwidth]{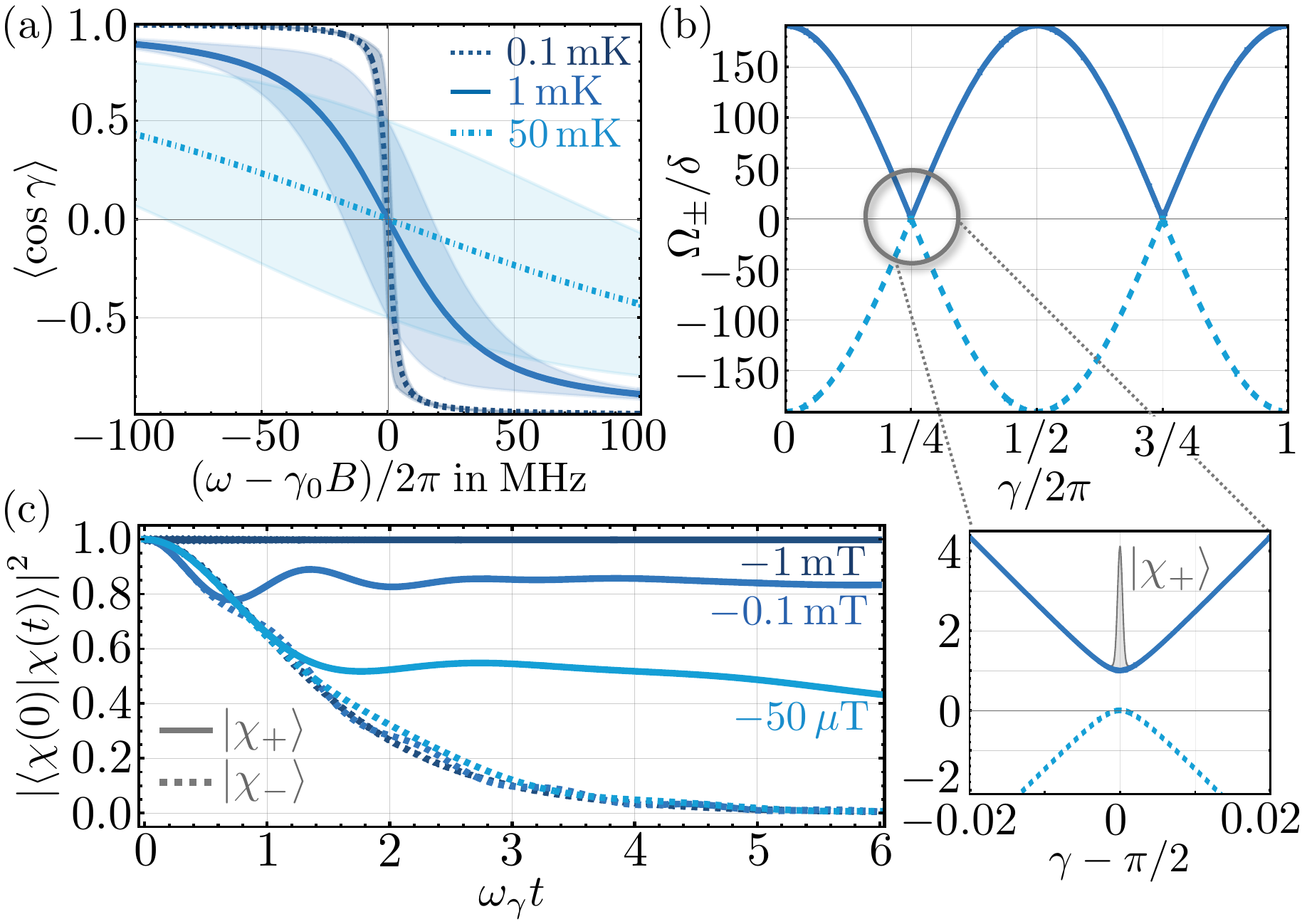}
\caption{(a) Steady-state alignment $\langle \cos \gamma \rangle$ as a function of the applied magnetic field. The variance is indicated by the shaded region. (b) Potential energy surfaces $\Omega_+$ (solid) and $\Omega_-$ (dashed) for $B=-0.5\,\mathrm{mT}$ and $\omega/2\pi=1\,\mathrm{MHz}$. The inset shows the avoided crossing around $\gamma=\pi/2$ where the spin-rotor state $|\chi_+\rangle$ can be stably trapped in the $|\sigma_x = +1\rangle$ spin state. (c) Spin transition probability for an initial state in spin  $|\sigma_x = +1\rangle$ (solid) and spin $|\sigma_x = -1\rangle$ (dashed) as a function of $\omega_\gamma t$ for different magnetic field strengths. The nanodiamond has an ellipsoidal shape with semiaxes $l_3 = 200$\,nm and $l_1 = l_2 = 0.3 l_3$.}
\label{fig:results1}
\end{figure}
\begin{align}\label{eq:hn}
    H_n=& \dfrac{p_\gamma^2}{2I_{\rm eff}} +  \dfrac{\gamma_0 B }{I\omega} S_3 p_\gamma + \dfrac{D_\mathrm{nv}}{2\hbar}(S_1^2 + S_2^2) \nonumber \\
    & + \frac{1}{2} (\omega-\gamma_0 B) \left [f_n (S_1 - i S_2) e^{-i \gamma} + {\rm h.c.}\right ] \nonumber \\
    & + \dfrac{D_\mathrm{nv}}{4\hbar} \left [ g_n (S_1+iS_2)^2 +{\rm h.c.} \right ].
\end{align}
This Hamiltonian describes the coupling between the NV spin and $\gamma$ rotations with effective inertia $I_{\rm eff} = I I_3/(I - I_3)$. The final two terms are the adiabatic potential (or potential energy surface) for fixed libration state $|n\rangle$, containing the magnetic potential as well as spin-rotation coupling and the Barnett effect on the NV zero-field splitting. The adiabatic terms are weighted by the overlap integrals
\begin{subequations}\label{eq:overlaps}
\begin{align}
f_n = & \langle n | \exp \left (-\frac{i p_\xi}{\hbar}  \frac{\hbar (\omega - \gamma_0 B)}{I \omega^2} \right )  | n \rangle, \\
g_n = & \langle n | \exp \left (-\frac{i p_\xi}{\hbar}  \frac{2 \hbar \gamma_0 B}{I \omega^2} \right )  | n \rangle,
\end{align}
\end{subequations}
for transitions between  displaced and nondisplaced libration states. Note that $f_n \simeq 1$ for $n \ll I^2 \omega^4/ \hbar^2 \gamma_0^2 B^2$ and likewise $g_n \simeq 1$. Given that $\hbar \gamma_0 B / I \omega^2 \ll 1$ for realistic magnetic fields and rotation rates, the Hamiltonian \eqref{eq:gyroham} becomes approximately independent of $n$. Thus, all adiabatic spin-rotor states $|\chi_n\rangle$ evolve according to the same effective Hamiltonian: 
\begin{align}\label{eq:adiabaticham}
    H_\mathrm{eff}= &\dfrac{p_\gamma^2}{2I_{\rm eff}}  +\dfrac{D_\mathrm{nv}}{\hbar}S_1^2 + \dfrac{\gamma_0 B }{\omega} \frac{S_3 p_\gamma}{I}\nonumber\\
    & +(\omega-\gamma_0 B)\left(S_1\cos\gamma-S_2\sin\gamma\right).
\end{align}
This Hamiltonian is the main result of this work and the starting point for the proposals discussed below. It accounts for spin-rotation coupling in three different ways. First, the planar $\gamma$ rotation occurs with the effective moment of inertia $I_{\rm eff}$ because $\xi$ librations are gyroscopically stabilized and adiabatically eliminated. Second, the rotation rate $\omega$ of the particle induces the effective spin-dependent potential in the  second line of the Hamiltonian \eqref{eq:adiabaticham}. Third, the $\mathbf{B}$ field induces a coupling of the spin to rotations around the symmetry axis of the particle as described by the last term in the first line. Note that this coupling is typically small and will be neglected in the following (see App.~\ref{app:adiabaticapproximation}). Next, we will study the implications of gyroscopic stabilization for the experimental observation of strong spin-rotation coupling.

{\it (a) Barnett splitting---} The Hamiltonian shows how the degeneracy of the $|S_1 = \pm \hbar\rangle$ states is lifted by the Barnett effect for $B = 0$. Specifically, neglecting $\gamma$ rotations by setting $\gamma = 0$ and $p_\gamma = 0$, the $|S_1 = \pm \hbar \rangle$ levels are split by the gyroscopic rotation frequency $\omega$. Note that the resonance frequency is also Doppler shifted by the particle rotation rate since the microwave drive is applied in the lab frame rather than in the rotating frame of the NV center \cite{chudo2015rotational,wood2018quantum,wood2020observation}.

{\it (b) Barnett alignment---} The described setup can be used to observe that the Barnett field $-\omega/\gamma_0$ tends to align prolate particles with their own rotation axis \cite{ma2021torque}. This follows from considering the hard magnet limit $|\omega - \gamma_0 B| \ll D_{\rm nv}$, where the spin is rigidly attached to the body frame $S_1 = \hbar m$ with $m \in\{-1,0,1\}$ and $S_2 = S_3 = 0$. Then the adiabatic Hamiltonian \eqref{eq:adiabaticham} implies that the particle revolves according to the spin-dependent potential $V_m(\gamma) = \hbar m (\omega - \gamma_0 B) \cos\gamma$ [see Fig.~\ref{fig:system}(c)], 
\begin{equation}
I_{\rm eff}\ddot{\gamma} = \hbar m (\omega - \gamma_0 B) \sin \gamma.
\end{equation}
For $B = 0$ and $m = 1$, the rotor tends to orient with $\gamma = \pi$ so that the rotation axis aligns with the spin axis. A finite magnetic field $B > 0$ weakens the impact of the Barnett field and exactly compensates it at $B = \omega/\gamma_0$, where the particle evolves freely in $\gamma$, as would be expected for an object carrying no spin degrees of freedom. Experimentally observing the alignment of the particle spin axis with the rotation axis is thus a unique signature of strong spin-rotation coupling in a rotating nanoparticle. It can be observed in state-of-the-art experiments with nanodiamonds spinning in quadrupole ion traps. This is illustrated in Fig.~\ref{fig:results1}(a), which shows the steady-state alignment $\langle \cos \gamma \rangle$ of the NV axis as a function of the rotation rate $\omega$ for three different temperatures. Note that a robust implementation in the presence of rotor asymmetries uses nanodiamonds hosting multiple NV centers (see App.~\ref{app:imperfections}).

{\it (c) Stabilization of spin superpositions---} For moderate magnetic fields, the superposition of the magnetic spin states $|\sigma_x=+1\rangle = (|S_1 = +\hbar\rangle + |S_1 = -\hbar\rangle)/\sqrt{2}$ becomes stabilized when the spin quantization axis lies in the rotation plane. Specifically, for $|\omega - \gamma_0 B| \lesssim D_{\rm nv}$, the nonmagnetic state $|S_1 = 0\rangle$ remains far detuned from the magnetic states $|S_1 = \pm \hbar\rangle$ for all orientations $\gamma$. The former can thus be adiabatically eliminated, yielding the spin-rotor Hamiltonian in the magnetic spin subspace
\begin{align}\label{eq:magham}
    H_\mathrm{mag} & = \dfrac{p_\gamma^2}{2 I_\mathrm{eff}}+\frac{\hbar \delta}{2} (\mathbb{1}+\sigma_x)\sin^2\gamma +\hbar g\sigma_z \cos\gamma.
\end{align}
Here, $\delta = (\omega - \gamma_0B)^2/D_{\rm nv}$, $g = (\omega - \gamma_0B)$, and $\sigma_z$ and $\sigma_x$ are Pauli matrices in the magnetic subspace of the NV spin. In contrast to the hard magnet limit, this Hamiltonian accounts for spin transitions as described by $\delta \neq 0$. Diagonalizing the spin-dependent part of Eq.~\eqref{eq:magham} yields the potential energy surfaces $\hbar \Omega_\pm(\gamma)/2$ for the $\gamma$ dynamics
\begin{equation}\label{eq:energy_surface}
\Omega_\pm(\gamma) = \delta \sin^2 \gamma \pm \sqrt{\delta^2 \sin^4 \gamma + 4g^2 \cos^2\gamma},
\end{equation}
as depicted in Fig.~\ref{fig:results1}(b). They exhibit an avoided crossing at $\gamma = \pi/2$ (and $3\pi/2$), in addition to the potential minima at $\gamma = 0, \pi$ for $|S_1 = \pm \hbar\rangle$ already observed in the hard magnet limit. The energies can be linearized near the avoided crossing yielding $\hbar\Omega_\pm(\gamma) = \pm \hbar \delta/2 \pm \hbar g^2 (\gamma - \pi/2)^2/\delta$, showing that the spin rotor is harmonically trapped in the spin state $|\sigma_x=+1\rangle$. The ground state of the resulting harmonic potential is stably trapped for $|\delta | \gg |g \sigma_\gamma|$, where $\sigma_\gamma = (\hbar \delta / 8 I_{\rm eff} g^2)^{1/4}$ is the ground state width in the effective harmonic potential (see App.~\ref{app:spintransition}). The state becomes unstable if this condition is violated, as illustrated in Fig.~\ref{fig:results1}(c) for three different magnetic field strengths, or when the spin superposition decoheres. 

{\it (d) Spin-controlled quantum interference---} For strong magnetic fields, $|\omega - \gamma_0 B| \simeq D_{\rm nv}$, the quantum state of the NV spin can be used to generate and read out superpositions of the angle $\gamma$. Specifically, the magnetic state $|S_1 = +\hbar \rangle$ can be adiabatically eliminated for small librations around $\gamma \simeq 0$, restricting the spin to the two levels $\left | \uparrow \right\rangle = |S_1 = - \hbar\rangle$ and $\left | \downarrow \right \rangle = |S_1 = 0\rangle$ with detuning $\Delta = D_{\rm nv} - g$. For small amplitudes and detunings $ \langle \gamma^2\rangle  \ll  |\Delta|/|g| \ll 1$, the two spins are effectively decoupled and the resulting dispersive spin-rotor Hamiltonian follows as (App.~\ref{app:interference})
\begin{equation}
    H_{\rm d} = \frac{p_\gamma^2}{2 I_\mathrm{eff}}+\dfrac{\hbar g}{8} \gamma^2+\dfrac{\hbar \Delta}{2} \sigma_z + \frac{3\hbar g}{8}\left(1 + \frac{4g}{3\Delta}\right)\gamma^2\sigma_z.
    \label{eq:H_dispersive}
\end{equation}
\begin{figure}[t!]
\includegraphics[width=\columnwidth]{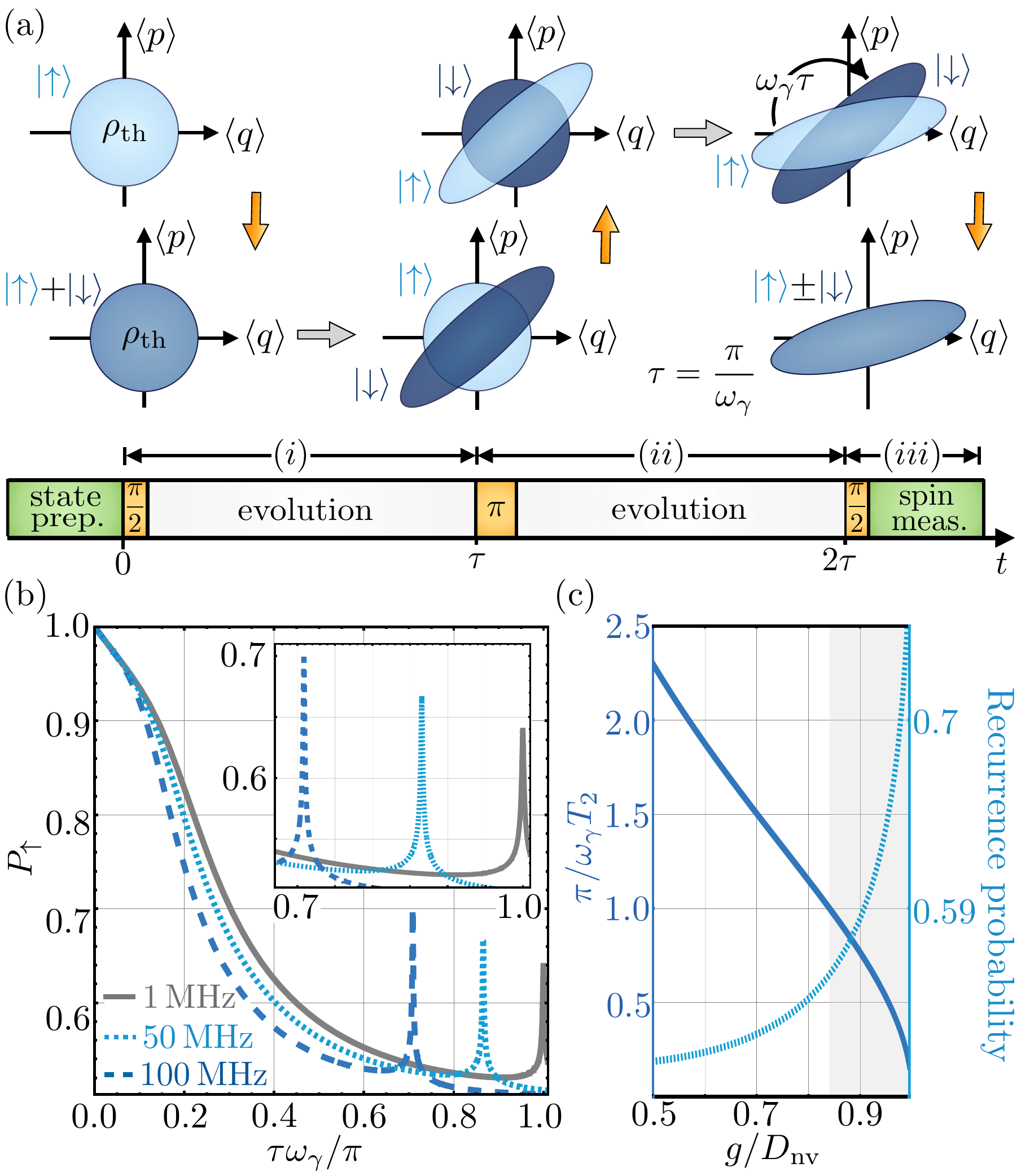}
\caption{(a) Pulse sequence and state evolution of the interference protocol. (b) Probability to measure $\left|\uparrow\right>$ at the end of the interferometer sequence as a function of interferometer time $\tau\omega_\gamma/\pi$ (evaluated at $\omega/2\pi=1\,\mathrm{MHz}$) for $B=-95\,\mathrm{mT}$ and different rotation frequencies $\omega/2\pi=1\,\mathrm{MHz}$ (gray, solid), $50\,\mathrm{MHz}$ (light blue, dotted), $100\,\mathrm{MHz}$ (dark blue, dashed). The inset shows the probability close to rephasing. (c) Duration  of the full protocol (dark blue, solid) and recurrence probability (light blue, dotted) depending on $g/D_\mathrm{nv}$. The shaded region indicates when the interferometer time is exceeded by the $T_2$ time of the NV spin. The nanodiamond has an ellipsoidal shape with semiaxes $l_3 = 100$\,nm and $l_1 = l_2 = 0.2 l_3$.}
\label{fig:protocol}
\end{figure}
This dispersive Hamiltonian~\eqref{eq:H_dispersive} can be used to create and read out nonclassical states of the rotor with a simplified version of the scheme discussed in Ref.~\cite{rusconi2022spin}. Specifically, for $\Delta>0$, the spin is initialized in $\left|\uparrow\right\rangle$, while the rotor is prepared in a thermal oscillator state $\rho_\mathrm{th}$ with corresponding frequency $\omega_{\gamma} = \sqrt{\hbar g(1+g/\Delta)/I_\mathrm{eff}}$ and temperature $T$, $\rho=\rho_\mathrm{th}\otimes\left|\uparrow\right>\!\left<\uparrow\right|$. The interference protocol is then implemented in three steps [see  Fig.~\ref{fig:protocol}(a)]: (i) a microwave $\pi/2$-pulse is applied to prepare the spin in a superposition $\left | \uparrow \right \rangle \to (\left | \uparrow \right \rangle + \left | \downarrow \right \rangle)/\sqrt{2}$. The oscillator then evolves for a time $\tau$, during which the branch with spin $\left|\downarrow\right \rangle$ becomes squeezed. (ii) At time $t=\tau$, a microwave $\pi$-pulse reverses the role of the spin $\left|\uparrow\!(\downarrow)\right>\rightarrow\left|\downarrow\!(\uparrow)\right>$, followed by a second evolution over a time $\tau$ which now squeezes the thermal oscillator in the second spin branch. (iii) A final $\pi/2$-pulse at $t=2\tau$ closes the interferometer through $\left|\uparrow\!(\downarrow)\right>\rightarrow(\left|\uparrow\right>\pm\left|\downarrow\right>)/\sqrt{2}$, creating a superposition of squeezed states. The microwave pulses are applied in the lab frame and therefore experience a rotational Doppler shift in the NV frame; however, the drive remains resonant throughout the rotation, ensuring coherent spin control during the entire protocol (see App.~\ref{app:doppler}).  The evolution of the oscillator during the pulse is negligible as the pulse duration is much shorter than the timescale of its dynamics. The probability of measuring spin $\left | \uparrow\!(\downarrow)\right >$ after the interferometer sequence is $ P_{\uparrow\downarrow}=1/2\pm\exp(-2\tau/T_2)\sqrt{|\lambda_\tau|}\cos[\mathrm{arg}(\lambda_\tau)/2]/2$, with the Hahn-echo spin dephasing time $T_2$ and the time-dependent complex number $\lambda_\tau$ (see App.~\ref{app:interference}). For small temperatures $T \ll \hbar \omega_\gamma/k_{\rm B}$, with the Boltzmann constant $k_\mathrm{B}$, it is determined by
\begin{align}
   \frac{1}{\lambda_\tau}=1+(1-\mathrm{e}^{2i\omega_\gamma\tau})\sinh^2 \left (\frac{\sqrt{\hbar} g \tau}{\sqrt{I_{\rm eff} \Delta}}\right ).
   \label{eq:probability}
\end{align} 
The spin probability rephases for $\tau=\pi/\omega_\gamma$, where $\lambda_\tau = 1$, due to a perfect overlap of the states of the two branches, reduced only by NV spin decoherence with the realistic dephasing time of $T_2=10\,\mu\mathrm{s}$ \cite{barry2020,chen2023extending,du2024single}. Other environmental decoherence channels remain negligible in the regime considered (see App.~\ref{app:decoherence}). This can be seen in Fig.~\ref{fig:protocol}(b), which demonstrates that increasing the  rotation frequency $\omega$ enhances the interference signal by reducing the total interferometer time. Figure~\ref{fig:protocol}(c) shows the total interferometer time and the recurrence probability $P_\uparrow(\tau=\pi/\omega_\gamma)$ as a function of the spin-rotor coupling $g/D_{\rm nv}$. The shaded region indicates where the NV's $T_2$ time exceeds the interferometer time, resulting in a significant enhancement of the interference signal.

The proposed experiments for gyroscopically stabilized spin rotors are within reach given the recent achievements in the field of levitated particles \cite{delord2020spin,delic2020cooling,magrini2021real,tebbenjohanns2021quantum,jin2024quantum,perdriat2024rotational}. Electrically charged nanodiamonds hosting NV centers have been trapped, manipulated, and monitored both via light scattering as well as by preparation and measurements of the embedded spins (state initialization with a green laser and microwave driving to the magnetic states~\cite{segawa2023nanoscale}) \cite{delord2018ramsey, delord2017strong,hoang2016electron}. Gyroscopic spinning of charged particles with embedded spin degrees of freedom was recently demonstrated \cite{jin2024quantum,perdriat2024rotational}, achieving MHz rotation rates. Illuminating the particle with circularly polarized light can further boost the rotation rate beyond GHz \cite{reimann2018ghz,ahn2018optically,jin2021}. 

In summary, we discussed the gyroscopic stabilization of rapidly spinning nanodiamonds with a single embedded NV spin, leading to an effective spin-rotor Hamiltonian containing only a single rotational degree of freedom. We proposed several schemes to use this Hamiltonian to probe spin-rotation coupling in state-of-the-art setups and to perform  mechanical quantum interference experiments. Our work presents the first steps toward applications of gyroscopically stabilized spin rotors.  

{\it Acknowledgments---} We thank Cosimo C. Rusconi, Sanchar Sharma, Maxime Perdriat, Julen Pedernales, and Klaus Hornberger for stimulating discussions and comments on the manuscript. V.W. and B.A.S. acknowledge funding by the Carl-Zeiss-Foundation through the Project QPhoton. V.W. acknowledges support by the Deutsche Forschungsgemeinschaft (DFG, German Research Foundation) via the Research Training Group 1995 ‘Quantum Many-Body Methods in Condensed Matter Systems’ Project No. 240766775. V.W. and S.V.K. are supported by the German Federal Ministry of Research, Technology and Space (BMFTR) Project QECHQS under Grant No. 16KIS1590K. B.A.S. is supported by the DFG--510794108.

\appendix

\section{Rotation tensor and spin-rotor Hamiltonian}\label{app:body-frame}

The rotation tensor $\mathrm{R}(\Omega)$ rotates the laboratory frame axes $\mathbf{e}_x,\mathbf{e}_y,\mathbf{e}_z$ into the body-fixed principal axes $\mathbf{n}_1,\mathbf{n}_2,\mathbf{n}_3$
\begin{equation}
    \mathbf{n}_\mu(\Omega)=\mathrm{R}(\Omega)\mathbf{e}_j.
\label{eq:trafo_lab-body}
\end{equation}
Choosing Euler angles $\Omega=(\alpha,\beta,\gamma)^T$ according to the \textit{z}-\textit{y}'-\textit{z}'' convention, we have
\begin{align}
\mathrm{R}(\Omega) = &\mathrm{R}_{z}(\alpha)\mathrm{R}_{y}(\beta)\mathrm{R}_{z}(\gamma) =
    \begin{pmatrix}
        \cos\alpha & -\sin\alpha & 0\\
        \sin\alpha & \cos\alpha & 0\\
        0 & 0 & 1
    \end{pmatrix}\nonumber\\
     &\begin{pmatrix}
        \cos\beta & 0 & \sin\beta \\
         0 & 1 & 0\\
        -\sin\beta & 0 & \cos\beta
    \end{pmatrix}
    \begin{pmatrix}
        \cos\gamma & -\sin\gamma & 0\\
        \sin\gamma & \cos\gamma & 0\\
        0 & 0 & 1
    \end{pmatrix}.
    \label{eq:rotationMatrix}
\end{align}
This implies that the principal axes in terms of the Euler angles read
\begin{subequations}
\begin{align}
    \mathbf{n}_1&=\begin{pmatrix}
        \cos\alpha\cos\beta\cos\gamma-\sin\alpha\sin\gamma\\ 
        \sin\alpha\cos\beta\cos\gamma+\cos\alpha\sin\gamma \\
        -\sin\beta\cos\gamma
    \end{pmatrix},\\
    \mathbf{n}_2&=\begin{pmatrix}
        -\cos\alpha\cos\beta\sin\gamma-\sin\alpha\cos\gamma\\
        -\sin\alpha\cos\beta\sin\gamma+\cos\alpha\cos\gamma\\
        \sin\beta\sin\gamma
    \end{pmatrix},\\
    \mathbf{n}_3&=\begin{pmatrix}
    \cos\alpha\sin\beta\\
   \sin\alpha \sin\beta\\
    \cos\beta
    \end{pmatrix}.
\end{align}
    \label{eq:labAxes_bodyFrame}
\end{subequations}
The Hamiltonian of a rigid rotor with embedded spin in an external magnetic field reads \cite{rusconi2022spin}
\begin{equation}
    H_0= \dfrac{1}{2}(\mathbf{J}-\mathbf{S})\cdot \mathrm{I}^{-1}(\mathbf{J}-\mathbf{S}) + \dfrac{D_\mathrm{nv}}{\hbar}(\mathbf{n}\cdot\mathbf{S})^2  +\gamma_0\mathbf{B}\cdot\mathbf{S}.
    \label{eq:H0}
\end{equation}
The first term is the rotational kinetic energy with $\mathbf{J} - {\bf S}$ the kinetic angular momentum vector operator of the particle. The second term describes the preferred alignment of the NV quantization axis with the body-fixed ${\bf n}(\Omega)$ axis. These two terms conserve the total angular momentum ${\bf J}$ so that spin and rotation are effectively coupled as witnessed in the Einstein-de Haas and Barnett effects \cite{einstein1915experimental,barnett1915magnetization}. The third term describes both the torque exerted by the magnetic field on the mechanical rotation as well as the orientation-dependent precession of the NV spin in the external field. For instance, the energy of the spin eigenstates $|S_{\bf n} = \pm1\rangle$ of the spin operator $S_{\bf n} = {\bf n}(\Omega)\cdot {\bf S}$ is Zeeman-split by the magnetic field by the orientation-dependent amount $2 \hbar \gamma_0 {\bf B} \cdot {\bf n}(\Omega)$.

The components of the total and the spin angular momentum operator follow the canonical angular momentum commutation relations. Denoting the body-frame components with Greek indices, $J_\mu= \mathbf{J}\cdot \mathbf{n}_\mu(\Omega)$ and $S_\mu= \mathbf{S}\cdot \mathbf{n}_\mu(\Omega)$, with $\mu=1,2,3$, these commutation relations are \cite{vanvleck1951}
\begin{subequations}
\begin{equation}
[J_\lambda,J_\mu]=-i\hbar\varepsilon_{\lambda\mu\nu}J_\nu,
\quad{\rm and} \quad
    [S_\lambda,S_\mu]=i\hbar\varepsilon_{\lambda\mu\nu}S_\nu,
\end{equation}
as well as
\begin{equation}
    [J_\lambda,S_\mu]=0.
\end{equation}
\end{subequations}

The components of the total and spin angular momentum vector operators can also be given in the lab frame, $J_j= \mathbf{J}\cdot \mathbf{e}_j$ and $S_j = {\bf S}\cdot {\bf e}_j$. Then $J_j$ and $S_k$ do not commute as the corresponding canonical commutation relations are
\begin{subequations}
\begin{equation}
[J_k,J_\ell]=i\hbar\varepsilon_{k \ell m}J_m \quad{\rm and}\quad   [S_k,S_\ell]=i\hbar\varepsilon_{k \ell m}S_m,
\end{equation}
as well as
\begin{equation}
[J_k,S_\ell]=i\hbar\varepsilon_{k \ell m}S_m.
\end{equation}
\end{subequations}
Note that $[J_j,J_\mu] = 0$. Thus the operators $({\bf J}^2, J_3,J_z, {\bf S}^2, S_{\bf n})$ form a complete set of commuting observables \cite{rusconi2016magnetic}.

The canonical momentum operators $p_\alpha, p_\beta, p_\gamma$ are related to the angular momentum vector operator through the relations
\begin{subequations}
\begin{align}
    p_\alpha = & {\bf J}\cdot {\bf e}_z, \\
    p_\beta = & \frac{1}{2}  {\bf J}\cdot \left (-\sin\alpha {\bf e}_x + \cos \alpha {\bf e}_y \right ) + {\rm h.c.},  \\
    p_\gamma = & {\bf J} \cdot {\bf n}_3(\Omega).
\end{align}
\end{subequations}
This implies that the body-frame components of the angular momentum operator can be expressed through 
\begin{subequations}
\begin{align}
J_1&=p_\beta\sin\gamma-p_\alpha\csc\beta\cos\gamma+ \frac{\cot\beta}{2} \{p_\gamma,\cos\gamma \},\\
J_2&=p_\beta\cos\gamma + p_\alpha\csc\beta\sin\gamma- \frac{\cot\beta}{2} \{p_\gamma,\sin\gamma\},\\
 J_3&=p_\gamma.
\end{align}
\label{eq:app_J_components_body}
\end{subequations}
We emphasize that these canonical angular momentum operators differ from the purely kinetic angular momentum \(\mathbf{L}\) that appears in the configuration-space description based on the Euler-angle velocities \(\dot{\alpha},\dot{\beta},\dot{\gamma}\). 
The Hamiltonian used throughout the manuscript is formulated in terms of the canonical phase-space variables and therefore naturally incorporates the spin contribution to the total angular momentum \(\mathbf{J}=\mathbf{L}+\mathbf{S}\).

\section{Trapping potential in the rotating frame} \label{app:gyropotential}

The dynamics of the spin rotor is typically also affected by trapping torques used to control the nanoparticle motion. These torques do not directly affect the spin state and must be added to the Hamiltonian \eqref{eq:H0}. As the experimentally most relevant scenario, we consider the electric torques exerted by radio frequency electric quadrupole traps. Denoting the drive frequency by $\omega_{\rm ac}$ and the ac voltage by $U_{\rm ac}$, the electric field near the trap center is of quadrupole character,
\begin{equation}
    {\bf E}({\bf r},t) = \frac{U_{\rm ac}}{d^2_0} \cos(\omega_{\rm ac}t) \sum_{j = x,y,z} A_j ({\bf e}_j\cdot {\bf r}) {\bf e}_j,
\end{equation}
where the coefficients $A_j$, with $\sum_j A_j = 0$, determine the field gradient and $d_0$ is the characteristic length scale of the trap. Here, we aligned the lab frame with the main trapping axes of the quadrupole trap \cite{martinetz2021electric}. In practical situations the surface charge distribution of the ellipsoid is approximately body-fixed and homogeneous, so that the dipole moment can be neglected \cite{perdriat2024rotational} and there is no coupling between center-of-mass and rotational degrees of freedom \cite{martinetz2021electric}. The trapping torques are then characterized by the body-fixed quadrupole tensor
\begin{equation}
 \mathrm{Q}(\Omega)= \sum_{\mu = 1}^3 Q_{\mu}\mathbf{n}_\mu(\Omega)\otimes\mathbf{n}_\mu(\Omega)   ,
\end{equation}
which is symmetric and traceless $\sum_\mu Q_{\mu} = 0$. If the drive frequency is sufficiently fast in comparison to the particle dynamics, so that the small-amplitude micromotion can be temporally averaged, the large-amplitude macromotion is induced by the secular quadrupole potential \cite{martinetz2021electric}
\begin{align}
V_\mathrm{qu}(\Omega) = 
\dfrac{U_\mathrm{ac}^2}{36\omega_\mathrm{ac}^2 d_0^4}&\sum_{\lambda = 1}^3  \frac{1}{I_\lambda}  \left [\sum_{j = x,y,z} A_j \sum_{\mu,\nu = 1}^3 \varepsilon_{\mu \lambda \nu } Q_\mu \right. \nonumber \\
& \left. \vphantom{\sum_{\mu,\nu = 1}^3} \times [{\bf n}_\mu(\Omega) \cdot {\bf e}_j] [{\bf n}_\nu(\Omega)\cdot {\bf e}_j]\right ]^2.
\label{eq:V_eff_0}    
\end{align}
The quadrupole potential depends on all three Euler angles through the principal axes ${\bf n}_\mu(\Omega)$. The resulting trapping torques tend to align the principal axes of the particle ${\bf n}_\mu(\Omega)$ with the axes of the lab frame ${\bf e}_j$. Full three-dimensional alignment is possible if the particle has three distinct quadrupole moments $Q_1 \neq Q_2 \neq Q_3$ and if the trap is asymmetric $A_1\neq A_2 \neq A_3$. 

In the considered case of a symmetric particle and symmetric trap configuration, i.e. $A_1=A_2= -A_3/2$, the quadrupole potential reduces to a potential only confining the $\beta$-angle, namely
\begin{equation}
   V_\mathrm{qu}(\beta)= \dfrac{U_\mathrm{ac}^2(Q-Q_{3})^2}{16 I\omega_\mathrm{ac}^2d_0^4}\sin^2\beta \cos^2\beta,
   \label{eq:app_V_beta}
\end{equation}
and Eq.~\eqref{eq:H0} reduces to the Hamiltonian~\eqref{eq:hamiltonian}. For an electric trap with $U_\mathrm{ac}=2.5\,\mathrm{kV}$, $\omega_\mathrm{ac}/2\pi=0.5\,\mathrm{MHz}$, electrode spacing $
    d_0=350\,\mu\mathrm{m}$ \cite{martinetz2021electric}, particle dimensions $l_3 = 200\,\mathrm{nm}$, $l_1=l_2=0.3 l_3$, and surface charge density $\sigma=3.5\,\mu\mathrm{C}/\mathrm{m}^2$, the resulting harmonic frequency of the trapping potential in $\beta$ compared to a rotation frequency $\omega/2\pi=1\,\mathrm{MHz}$ is 
    \begin{equation}
        \omega_\beta=
  \sqrt{\frac{U_\mathrm{ac}^2(Q-Q_{3})^2}{8 I^2\omega_\mathrm{ac}^2d_0^4}}\approx 4\times 10^{-5}\, \omega,
    \end{equation}
    hence $\omega_\xi\simeq \omega$.
We move to a frame corotating with the rotor $U H U^\dagger+i\hbar\partial_t U U^\dagger$ with the unitary operator
 \begin{equation}
U(t)=\exp\left(i\dfrac{\omega t}{\hbar} p_\alpha-i\dfrac{I \omega}{\hbar}\alpha\right),
\label{eq:trafo_rotFrame2}
 \end{equation}
which adds the term $-\omega p_\alpha$ and replaces $p_\alpha$ by $p_\alpha + I \omega$ while leaving the remaining Hamiltonian unchanged. Using that in the rotating frame $p_\alpha=0$, the Hamiltonian reads
\begin{align}\label{eq:app_hamiltonian_rot0}
H_\mathrm{rot}=&\dfrac{p_\beta^2}{2I} +\dfrac{(I \omega - p_\gamma \cos\beta )^2}{2I\sin^2\beta}-\dfrac{p_\beta}{I}(S_1\sin\gamma+S_2\cos\gamma)\nonumber\\
 &+\left \{ \dfrac{I\omega- p_\gamma\cos\beta}{2I\sin^2\beta},S_1\cos\gamma-S_2\sin\gamma\right \} +V_{\rm qu}(\beta)\nonumber\\
 &-\gamma_0 B \left(S_1\sin\beta\cos\gamma -
    S_2\sin\beta\sin\gamma -
    S_3\cos\beta \right)\nonumber\\
&+\dfrac{S_1^2+S_2^2}{2I}+\dfrac{(p_\gamma-S_{3})^{2}}{2I_3}+\frac{D_\mathrm{nv}}{\hbar} S^2_1.
\end{align}
Here, a number of terms can be neglected: First, the terms $(S_1^2 + S_2^2)/2I$ and $S_3^2/2I_3$ are negligible in comparison to the NV zero-field splitting $D_{\rm nv}S_1^2/\hbar$ and can thus be dropped. Second, the spin-rotation coupling in $\beta$ as described by the last term in the first line is always suppressed compared to the $\beta$ trapping potential as well as to the gyroscopic potential in $\gamma$ and can thus be neglected. Third, the spin-rotation coupling in $\gamma$ as described by $-p_\gamma S_3/I_3$ as well as by the second line is always smaller than the gyroscopic coupling described by the second term in the first line. Considering $S_3=\hbar$ and small fluctuations in $\beta$ characterized by the zero-field amplitude $\xi_0=\sqrt{\hbar/2 I \omega}$, the spin-rotation coupling compares to the gyroscopic coupling $p_\gamma \omega\xi$ as $S_3/I_3 \omega\xi_0\approx 10^{-4}$ ($l_3 = 200\,\mathrm{nm}$, $l_1=l_2=0.3 l_3$, $\omega/2\pi=1\,\mathrm{MHz}$), and thus can  be dropped. The simplified Hamiltonian reads
\begin{align}\label{eq:app_hamiltonian_rot}
H_\mathrm{rot}=&\dfrac{p_\beta^2}{2I} +\dfrac{(I \omega - p_\gamma \cos\beta )^2}{2I\sin^2\beta}+\dfrac{p_\gamma^2}{2I_3}+\frac{D_\mathrm{nv}}{\hbar} S^2_1+V_{\rm qu}(\beta)\nonumber\\
&-\gamma_0 B \left(S_1\sin\beta\cos\gamma -
    S_2\sin\beta\sin\gamma -
    S_3\cos\beta \right)\nonumber\\
    & + \dfrac{\omega}{\sin^2\beta}(S_1\cos\gamma-S_2\sin\gamma).
\end{align}

By linearizing around $\beta=\pi/2$ and only keeping terms up to second order in $\xi=\pi/2-\beta$, i.e. $1/\sin^{2}\beta\approx 1+\xi^2$, $\cos\beta/\sin^{2}\beta\approx \xi$, $\cos^2\beta\sin^{2}\beta\approx \xi^2$, as well as neglecting the terms proportional to $S_1\xi^2$, $S_2\xi^2$, and $p_\gamma^2\xi^2$,  the Hamiltonian is
\begin{align}
H_{\rm rot} \approx &  \dfrac{p_\xi^2}{2I} + \dfrac{I}{2}\omega_\xi^2 \xi^2+(\gamma_0B S_3 - \omega p_\gamma)  \xi  + \frac{p^2_\gamma}{2I_3} +\frac{D_{\rm nv}}{\hbar}S_1^2\nonumber \\
&    + \left ( \omega - \gamma_0B \right )( S_1\cos\gamma -  S_2\sin \gamma ).
\end{align}
It can be seen that the particle's rotation adds the gyroscopic potential $I \omega^2 \xi^2/2$ to the quadrupole potential and the spin-rotation coupling induces an effective potential for $\gamma$.

\section{Adiabatic approximation} \label{app:adiabaticapproximation}
\begin{figure}[t!]
\includegraphics[width=0.9\columnwidth]{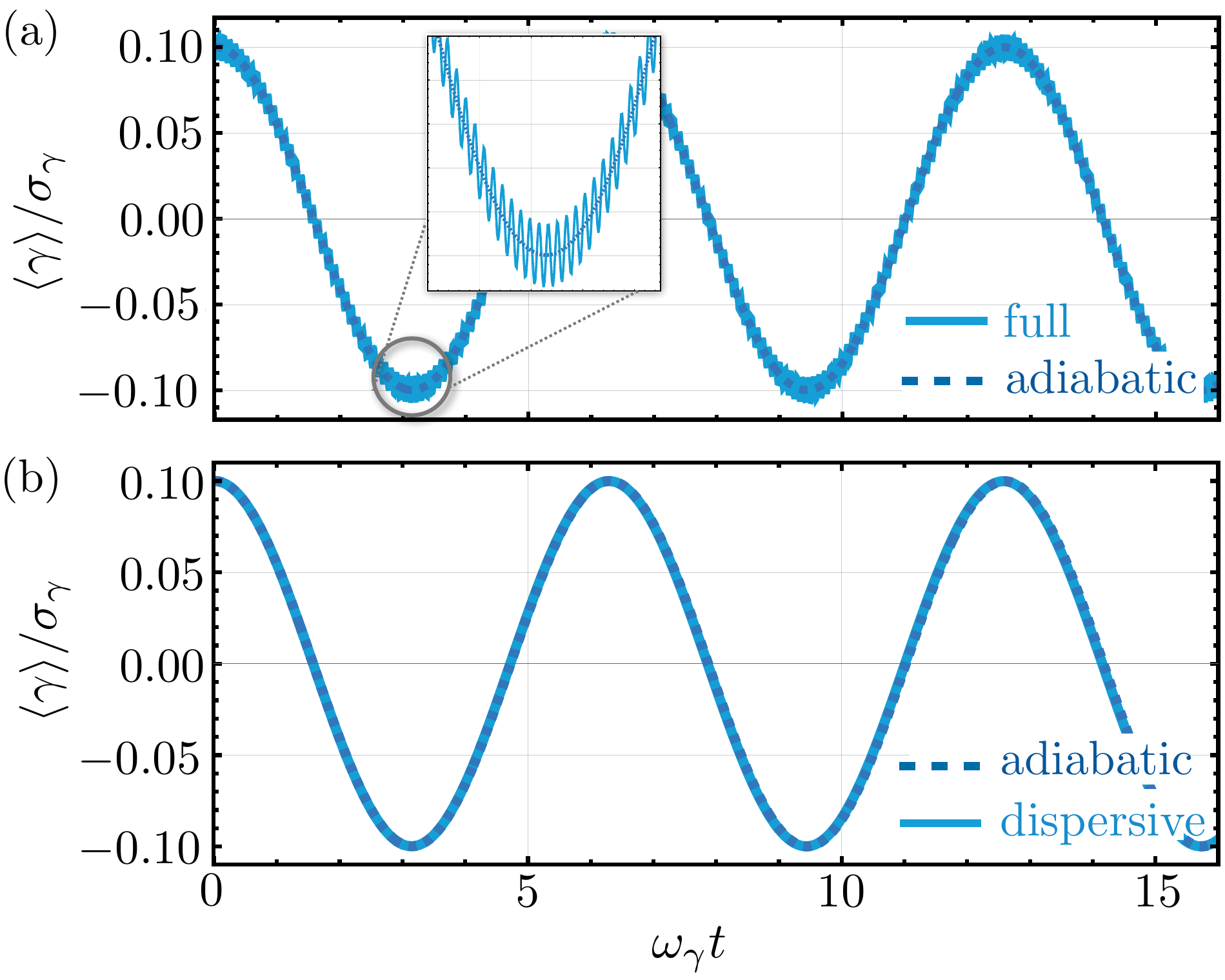}
\caption{Dynamics of $\gamma$ for the spin state $|S_1=-\hbar\rangle$ as a function of $\omega_0 t$, comparing the solution of (a) the gyroscopic Hamiltonian~\eqref{eq:gyroham} (light blue, solid) with the adiabatic Hamiltonian~\eqref{eq:adiabaticham} (dark blue, dashed) and (b) the adiabatic Hamiltonian~\eqref{eq:adiabaticham} with the dispersive Hamiltonian~\eqref{eq:H_dispersive} (light blue, solid). The nanodiamond has an ellipsoidal shape with semiaxes $l_3 = 200$\,nm and $l_1 = l_2 = 0.4 l_3$, $B=-102\,\mathrm{mT}$, $\omega/2\pi=0.1\,\mathrm{MHz}$, and $\sigma_\gamma=\sqrt{\hbar/2I_\mathrm{eff}\omega_\gamma}$. We approximated the Fock space as a finite space of dimension $d = 40$ and chose for both the initial states of the $\beta$- and $\gamma$-oscillator a coherent state $|\alpha_{\beta,\gamma}=0.1\rangle$.}
\label{fig:app_comparison1}
\end{figure}

The fast $\xi$ librations in Eq.~\eqref{eq:gyroham} can now be adiabatically eliminated by expanding an arbitrary spin-rotor state $|\phi\rangle$ in the oscillator eigenbasis $|n\rangle$ with $H_\xi|n\rangle=\hbar\omega (n+1/2)|n\rangle$ for $\omega_\xi \simeq \omega$. Using the ansatz \eqref{eq:stateexp}, 
\begin{equation}
 |\phi\rangle=  \sum_{n = 0}^\infty W \mathrm{e}^{-itH_\xi/\hbar}|n,\chi_n\rangle, 
\end{equation}
with $|n,\chi_n\rangle = |n\rangle\otimes |\chi_n\rangle$ and
\begin{equation}
    W=\exp \left(-\frac{ip_\xi}{\hbar} \frac{\omega p_\gamma - \gamma_0 B S_3}{I \omega^2}\right),
\end{equation}
the states $|\chi_n\rangle$ then evolve according to the coupled equations
\begin{align}\label{eq:coupledchannel}
 i\hbar\partial_t|\chi_n\rangle=& \left [ \frac{p_\gamma^2}{2I_{3}} -\hbar\omega\left(n+\dfrac{1}{2}\right) + h_{nn}\right ]|\chi_n\rangle \nonumber \\
 & +\sum_{m = 0 \atop {m \neq n}}^\infty e^{i \omega (n - m)t} h_{nm} |\chi_m\rangle.
\end{align}
Here we used that $[p_\gamma,W]=0$ and defined the coupling operators
\begin{align}
    h_{nm}&= \langle n|W^\dagger \left [H_\xi+(\gamma_0B S_3 -\omega p_\gamma)\xi +\dfrac{D_\mathrm{nv}}{\hbar}S_1^2 \right. \nonumber \\
    &\left. +(\omega-\gamma_0B)(S_1\cos\gamma-S_2\sin\gamma) \vphantom{\frac{D_{\rm nv}}{\hbar}}\right ] W|m\rangle.
\end{align}
Transitions between distinct libration states, as described by the second line of Eq.~\eqref{eq:coupledchannel}, can be neglected within the rotating wave approximation (Born-Oppenheimer approximation), so that only the diagonal coupling operators matter. A direct calculation shows that
\begin{align}
    h_{nn} = & \hbar\omega\left(n+\dfrac{1}{2}\right)-\dfrac{p_\gamma^2}{2I}-\dfrac{(\gamma_0 B S_3)^2}{2I\omega^2}+\dfrac{\gamma_0 B S_3 p_\gamma}{I\omega} \nonumber\\
    & + \langle n|V(\gamma,\mathbf{S})|n\rangle
\end{align}
with the spin-dependent potential
\begin{align}
V(\gamma,\mathbf{S})= &\dfrac{D_\mathrm{nv}}{\hbar}\left[S_1\cos\left(\dfrac{\gamma_0B }{I\omega^2}p_\xi\right)+S_2\sin\left(\dfrac{\gamma_0B }{I\omega^2}p_\xi\right)\right]^2\nonumber\\
&+(\omega-\gamma_0 B)\left[S_1\cos\left(\tilde{\gamma}\right)-S_2\sin\left(\tilde{\gamma}\right)\right].
\end{align}
Here the angle $\gamma$ is shifted by the coupling with $p_\xi$, i.e.
\begin{equation}
    \tilde{\gamma}=\gamma+\dfrac{p_\xi}{I\omega}-\dfrac{\gamma_0B p_\xi}{I\omega^2}.
\end{equation}
By expressing the cosine and sine functions in terms of exponential functions and defining the overlap integrals \eqref{eq:overlaps}, the effective Hamiltonian $H_n$ governing the dynamics of $|\chi_n\rangle$ can be written in the form \eqref{eq:hn}, where we neglected the term $\propto S_3^2$. Using experimentally accessible conditions ($l_3 = 200\,\mathrm{nm}$, $l_1=l_2=0.3 l_3$, $\omega/2\pi=1\,\mathrm{MHz}$, $T=300\,$K, $B=100\,\mathrm{mT}$), we find $n=k_\mathrm{B}T/\hbar\omega\approx 6.25\times 10^6$ such that $I^2 \omega^4/ \hbar^2 \gamma_0^2 B^2\approx3.8\times 10^{12} \gg n$, and likewise, the ratio $\hbar \gamma_0 B/I \omega^2 \approx 5\times 10^{-7}  \ll 1$. Thus, the overlap integrals can be approximately set to unity ($f_n,g_n\simeq1$), for which the effective Hamiltonian~\eqref{eq:adiabaticham} follows. In Fig.~\ref{fig:app_comparison1}(a) we compare the dynamics of $\gamma$ close to $\gamma=0$ for Eq.~\eqref{eq:gyroham} and Eq.~\eqref{eq:adiabaticham}. 

\subsection{Validity regime of the $\beta$-elimination}
\begin{figure}[t]
    \centering
    \includegraphics[width=0.65\columnwidth]{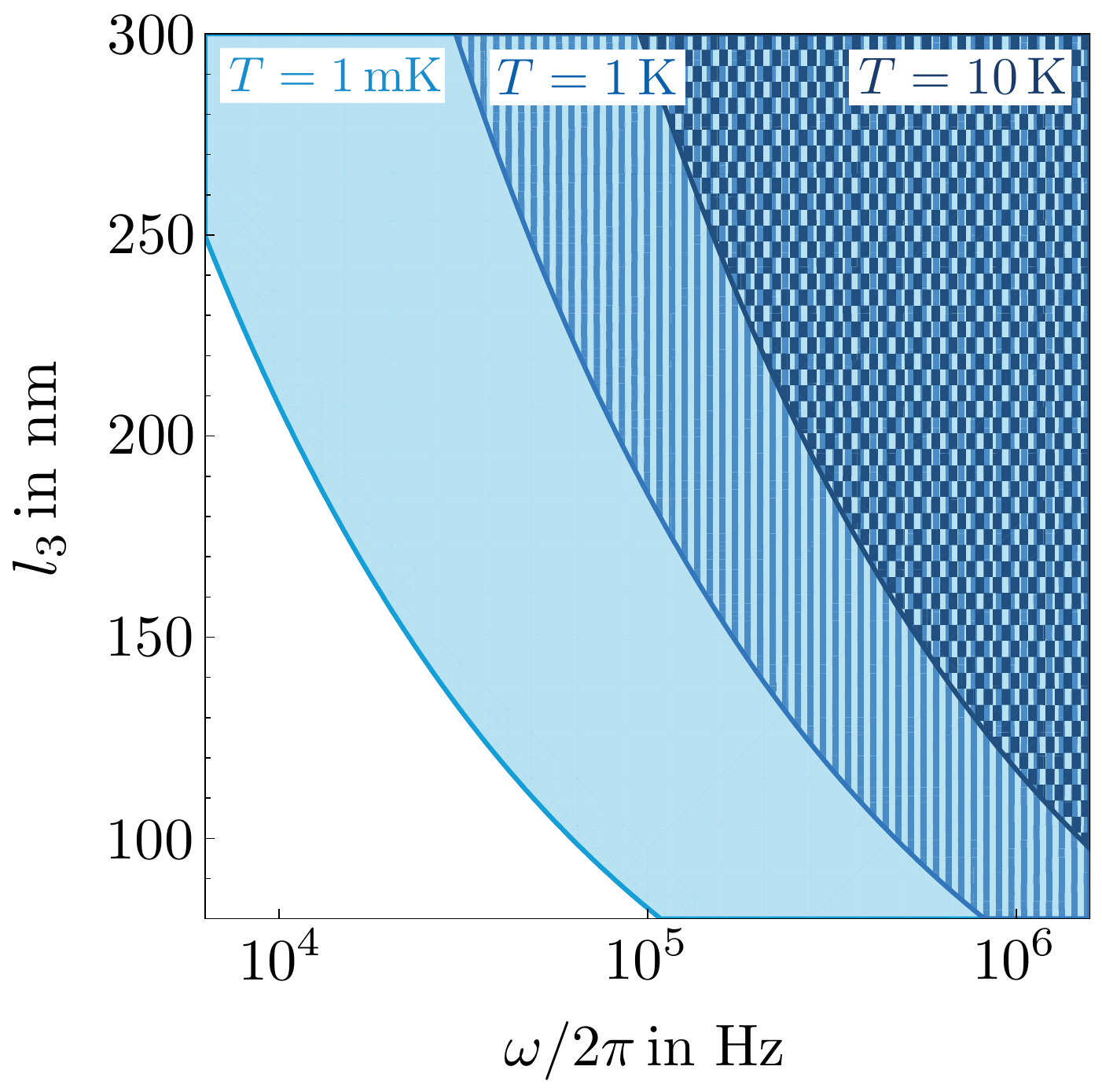}
    \caption{Validity of the adiabatic elimination of the $\xi$-mode. Parameter region satisfying Eq.~\eqref{eq:adiabatic_validity} as a function of the rotation frequency $\omega$ and the major semiaxis length $l_3$, with $l_1=0.2 l_3$, $B=-100\,\mathrm{mT}$ and different temperatures $T$ determining the mean occupation $n_\gamma$.}
    \label{fig:adiabatic_validity}
\end{figure}

The adiabatic elimination of the fast $\beta$ libration mode requires a well-defined separation of timescales between the $\beta$ and $\gamma$ degrees of freedom.  Quantitatively, the elimination requires
\begin{equation}
\left|\dfrac{p_\gamma}{I\omega}\right|\ll1, \quad \text{and}\quad \left|\dfrac{\hbar\gamma_0 B}{I\omega^2}\right|\ll 1,
\label{eq:adiabatic_validity}
\end{equation}
where the first condition suppresses transitions between distinct $\beta$ levels and ensures that the $\beta$-mode follows the slow $\gamma$-dynamics adiabatically, while the second inequality guarantees that the overlap integrals Eq.~\eqref{eq:overlaps} can be set to unity.
Using $p_\gamma=\sqrt{\hbar n_\gamma I_3 \omega_\gamma}$ with $n_\gamma$ the mean thermal occupation of $\gamma$ at temperature $T$, we quantitatively map the regime  where both criteria in Eq.~\eqref{eq:adiabatic_validity} are below 0.01. Fig.~\ref{fig:adiabatic_validity} shows the resulting parameter region as a function of the rotation frequency $\omega$ and particle size $l_3$ for $l_1=0.2 l_3$ and $B=-100\,\text{mT}$. Larger aspect ratios or weaker fields further further improve the separation of scales.

\subsection{Validity of neglecting the Zeeman-induced $S_3 p_\gamma$ coupling}
\begin{figure}[t!]
    \centering
    \includegraphics[width=\columnwidth]{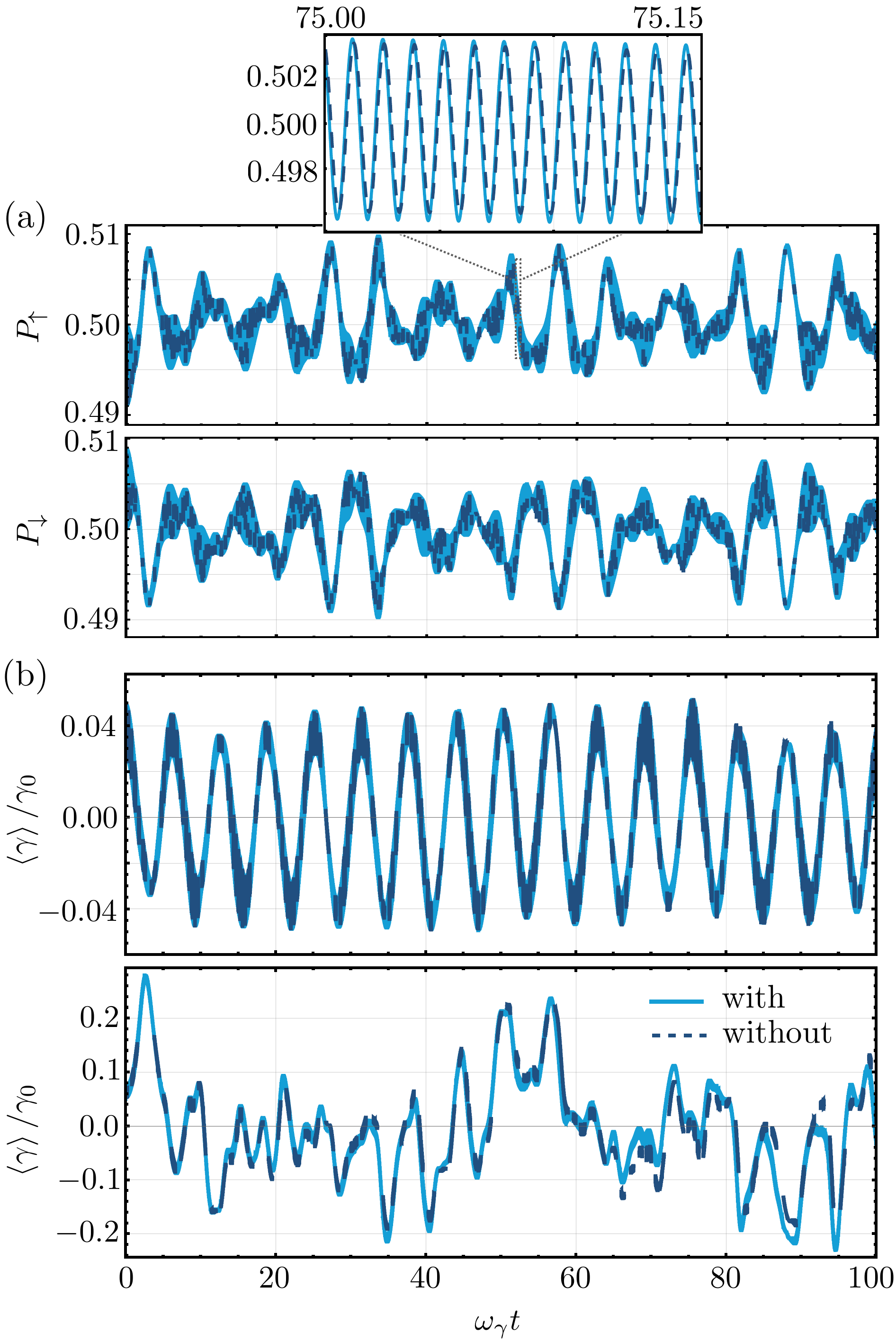}
\caption{Comparison of (a) $\gamma$- and (b) spin dynamics for the spin states $\left|S_1=-\hbar\right>=\left|\uparrow\right>$ (upper panel) and $\left|S_1=0\right>=\left|\downarrow\right>$ (lower panel), governed by the effective Hamiltonian $H_\mathrm{eff}$ given in Eq.~\eqref{eq:adiabaticham}. The system is initialized in the spin superposition  $(\left|\uparrow\right>+\left|\downarrow\right>)/\sqrt{2}$. Solid (light blue) curves include the Zeeman-induced spin-rotation coupling term $\gamma_0 B  S_3 p_\gamma/I\omega$, while dashed (dark blue) curves omit it. Results are shown for $l_3=100\,$nm, $B=-100\,$mT, $l_1=l_2=0.2 l_3$, and $\omega/2\pi=1\,\mathrm{MHz}$. The Fock space of the $\gamma$-oscillator is approximated as a finite space of dimension $d = 70$
and chose for the initial state
a coherent state $\left|\alpha_\gamma = 0.1\right>$.}
\label{fig:pgS3}
\end{figure}

To quantify the impact of the Zeeman-induced spin-rotation coupling term $\gamma_0 B S_3 p_\gamma/I\omega$, we compare Fig.~\ref{fig:pgS3} the full coupled $\gamma$-spin dynamics including this term [Eq.~\eqref{eq:adiabaticham}] with the resulting trajectories of those obtained from the simplified effective Hamiltonian in which it is omitted. The simulations were initialized in a spin superposition and evaluated for magnetic fields of $100\,$mT. For $\omega/2\pi\geq 1\,\mathrm{MHz}$ and $l_3=100\,$nm, the influence of the coupling term decreases rapidly, producing only minor quantitative deviations. For $l_3=200\,$nm the deviations already fall below numerical resolution. Thus, the Zeeman-induced coupling is strongly suppressed in the experimentally relevant regime. 

\section{Imperfections of the Rotor–Spin Model}\label{app:imperfections}
\subsection{NV-axis misalignment}
\begin{figure}[t!]
    \centering
    \includegraphics[width=\columnwidth]{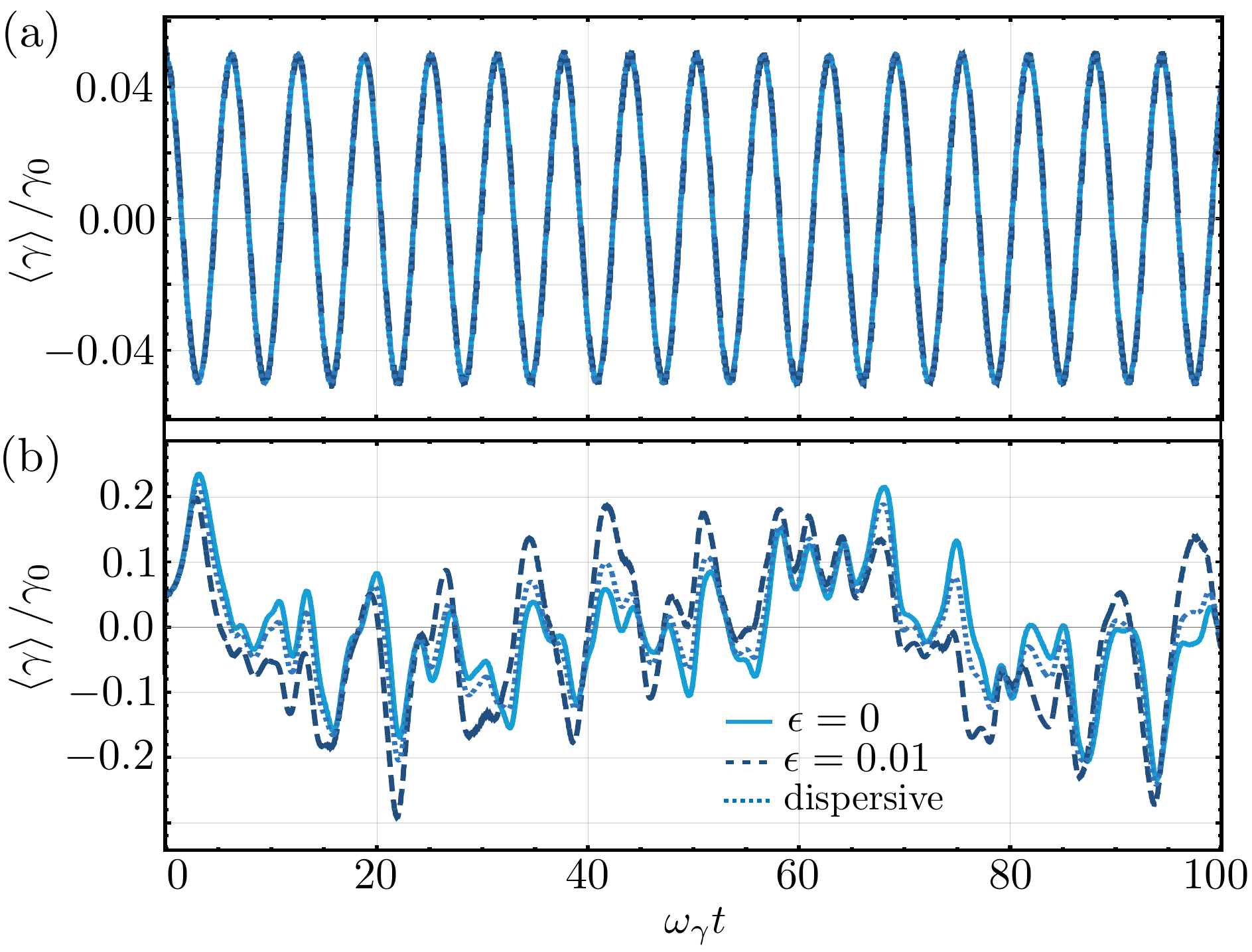}
\caption{Comparison of $\gamma$-dynamics for the spin states (a) $\left|S_1=-\hbar\right>=\left|\uparrow\right>$  and (b) $\left|S_1=0\right>=\left|\downarrow\right>$. The system is initialized in the spin superposition  $(\left|\uparrow\right>+\left|\downarrow\right>)/\sqrt{2}$. The solid (light blue) curve corresponds to $\mathbf{n}=\mathbf{n_1}$, while for the dashed (dark blue) curve includes a misalignment towards $\mathbf{n}=\mathbf{n_1}+\epsilon\mathbf{n}_3$ with $\epsilon=0.01$. The magnetic field is  $B=-55\,$mT, which still satisfies the dispersive conditions sufficiently as shown by the dotted line. The remaining parameters are $l_3=100\,$nm, $l_1=l_2=0.2 l_3$, and  $\omega/2\pi=1\,\mathrm{MHz}$. The Fock space of the $\gamma$-oscillator is approximated as a finite space of dimension $d = 80$ and chose for the initial state
a coherent state $\left|\alpha_\gamma = 0.1\right>$.}
\label{fig:misalign}
\end{figure}
A small tilt of the NV quantization axis relative to the principal inertia axis modifies the zero-field splitting term of the Hamiltonian. Considering the NV axis as ${\bf n} = {\bf n}_1 + \varepsilon {\bf n}_3$ with $\varepsilon\ll 1$,   the zero-field splitting becomes 
\begin{equation}
    H_\mathrm{zfs} = \dfrac{D_\mathrm{nv}}{\hbar} (S_1+\varepsilon S_3)^2\approx  \dfrac{D_\mathrm{nv}}{\hbar}\left(S_1^2+\varepsilon \{S_1,S_3\}\right),
\end{equation}
so that the effective Hamiltonian~\eqref{eq:adiabaticham} acquires the perturbation
\begin{align}
\delta H_\mathrm{eff}&= \dfrac{\varepsilon D_\mathrm{nv}}{\hbar} W^\dagger \{S_1,S_3\} W \nonumber\\
&=\dfrac{\varepsilon D_\mathrm{nv}}{2\hbar}g'_n\{S_1+i S_2,S_3\} +\mathrm{h.c.},
\end{align}
where $g'_n=\left<n\right|\mathrm{exp}\left(-i p_\xi\gamma_0 B/I\omega^2\right)\left|n\right>$. For the relevant regime, $g'_n\simeq 1$, the additional term reduces to
\begin{equation*}
\delta H_\mathrm{eff}=\dfrac{\varepsilon D_\mathrm{nv}}{\hbar}\{S_1,S_3\}.
\label{eq:eq:deltaHeff}
\end{equation*}
Restricting to the two-level subspace $\{\left|0\right>,\left|-\hbar\right>\}$, this perturbation mixes the spin components and rotates the eigenbasis by a small angle $\theta$ away from the $S_1$ eigenstates. The dressed eigenstates read
\begin{equation}
\left|\uparrow'\right>=\cos\theta\left|\uparrow\right>+i\sin\theta\left|\downarrow\right>,\quad  \left|\downarrow'\right>=-\cos\theta\left|\downarrow\right>-i\sin\theta\left|\uparrow\right>,
\end{equation}
with  mixing angle $\tan2\theta=\sqrt{2}\varepsilon D_\mathrm{nv}/|\Delta|$. The original basis $\left|\uparrow\right>$, $\left|\downarrow\right>$ remains a good approximation whenever
\begin{equation}
    \theta\simeq \dfrac{\varepsilon D_\mathrm{nv}}{\sqrt{2}\Delta }\ll 1\quad\Longleftrightarrow\quad |\Delta|\gg \dfrac{\varepsilon D_\mathrm{nv}}{\sqrt{2}}.
\end{equation}
 Numerical simulations (see Fig.~\ref{fig:misalign}) show, that for NV-axis deviations $\varepsilon\sim10^{-2}$, and for magnetic fields $|B|=55\,\mathrm{mT}$ in the dispersive regime, the induced corrections to the spin-rotation dynamics remain negligible on the interferometric timescale.

We note that precise control of NV orientation in nanodiamonds is routinely achievable with current fabrication and pre-selection techniques, ensuring that small misalignments are well within experimental tolerances \cite{perdriat2021spin,achard2020chemical,chen2017laser,kinouchi2023laser}.

\subsection{Shape asymmetry}
To assess the impact of small deviations from perfect axial symmetry, we extend the model to slightly asymmetric rotors with $I_2 = I_1 (1 - \delta_I)$, where $\delta_I\ll 1$. This asymmetry adds the kinetic-energy term an additional term 
$\delta_I(\mathbf{J}_2-\mathbf{S}_2)^2/2 I_1$, which weakly couples $\gamma$ to $S_2$ and generates a $\gamma$-dependent potential, while the angular momentum component $p_\alpha =\mathbf{J}\cdot\mathbf{e}_z$ remains conserved. Transforming to the corotating frame and linearizing in $\xi=\pi/2-\beta$, we obtain the correction to Eq.~\eqref{eq:gyroham},
\begin{align}
\delta H_{\rm rot} &=  \dfrac{\delta_I p_\xi^2}{2I_1}\cos^2\gamma + \dfrac{I_1  \delta_I\omega^2}{2} (1+\xi^2)\sin^2\gamma   -\delta_I\omega S_2 \sin\gamma \nonumber\\
&-\delta_I\omega p_\gamma \xi\sin^2\gamma+ \delta_I\omega p_\xi\sin\gamma\cos\gamma,
\end{align}
where we retain only the dominant term coupling $S_2$ to the rotational degrees of freedom. The leading effect is the rotation-induced potential $\propto\delta_I\omega^2\sin^2\gamma$, which modifies the $\gamma$ frequency. Importantly, this additional potential does not affect the interferometric protocol, which depends only on the frequency shift induced by the different spin configurations. 

For the Barnett-induced alignment along the rotation axis to remain valid, the asymmetry must satisfy 
\begin{equation}
\delta_I\ll \dfrac{\hbar(\omega-\gamma_0 B)}{I_1\omega^2}.
\end{equation}
For single NV centers this condition can be stringent, but in practice the effective Barnett alignment scales with the total magnetic moment. Nanodiamonds hosting multiple NV centers therefore relax this requirement, making the alignment condition easier to satisfy experimentally.

\section{Barnett anti-alignment}
We calculate the steady-state alignment $\langle\cos\gamma\rangle$ of the NV axis by considering a Boltzmann probability distribution of $\gamma$, i.e.
\begin{equation}
    \mathrm{prob}(\gamma)=\dfrac{1}{Z}\exp\left[-\dfrac{V_m(\gamma)}{k_\mathrm{B}T}\right].
\end{equation}
Here $k_\mathrm{B}$ denotes the Boltzmann constant and $Z$ the partition function given as
\begin{equation}
    Z=\int_0^{2\pi}\exp\left(-m \kappa \cos\gamma\right) = 2\pi I_0(m \kappa),
\end{equation}
where we defined $\kappa=\hbar (\omega-\gamma_0 B)/k_\mathrm{B}T$ and $I_\ell(\cdot)$ is the $\ell$-th order modified Bessel function of the first kind. The expectation value of $\cos\gamma$ follows as
\begin{align}
   \langle\cos\gamma\rangle &=-\dfrac{1}{Z}\dfrac{1}{m}\dfrac{\mathrm{\partial Z}}{\partial \kappa}=-\dfrac{I_1(\kappa m)}{ I_0(\kappa m)}.
\end{align}
Analogously, $\langle\cos^2\gamma\rangle = (\mathrm{\partial}^2Z/\partial \kappa^2) / Z m^2$ such that the variance is obtained as
\begin{equation}
    \mathrm{var}(\cos\gamma)=\frac{1}{2} + \dfrac{I_2(\kappa m)}{ 2I_0(\kappa m)}-\left[\dfrac{I_1(\kappa m)}{ I_0(\kappa m)}\right]^2,
\end{equation}
implying that $0\leq {\rm var}(\cos\gamma) \leq 1/2$. 

\section{Stabilization of spin superpositions}\label{app:spintransition}
\begin{figure}[t!]
\includegraphics[width=0.9\columnwidth]{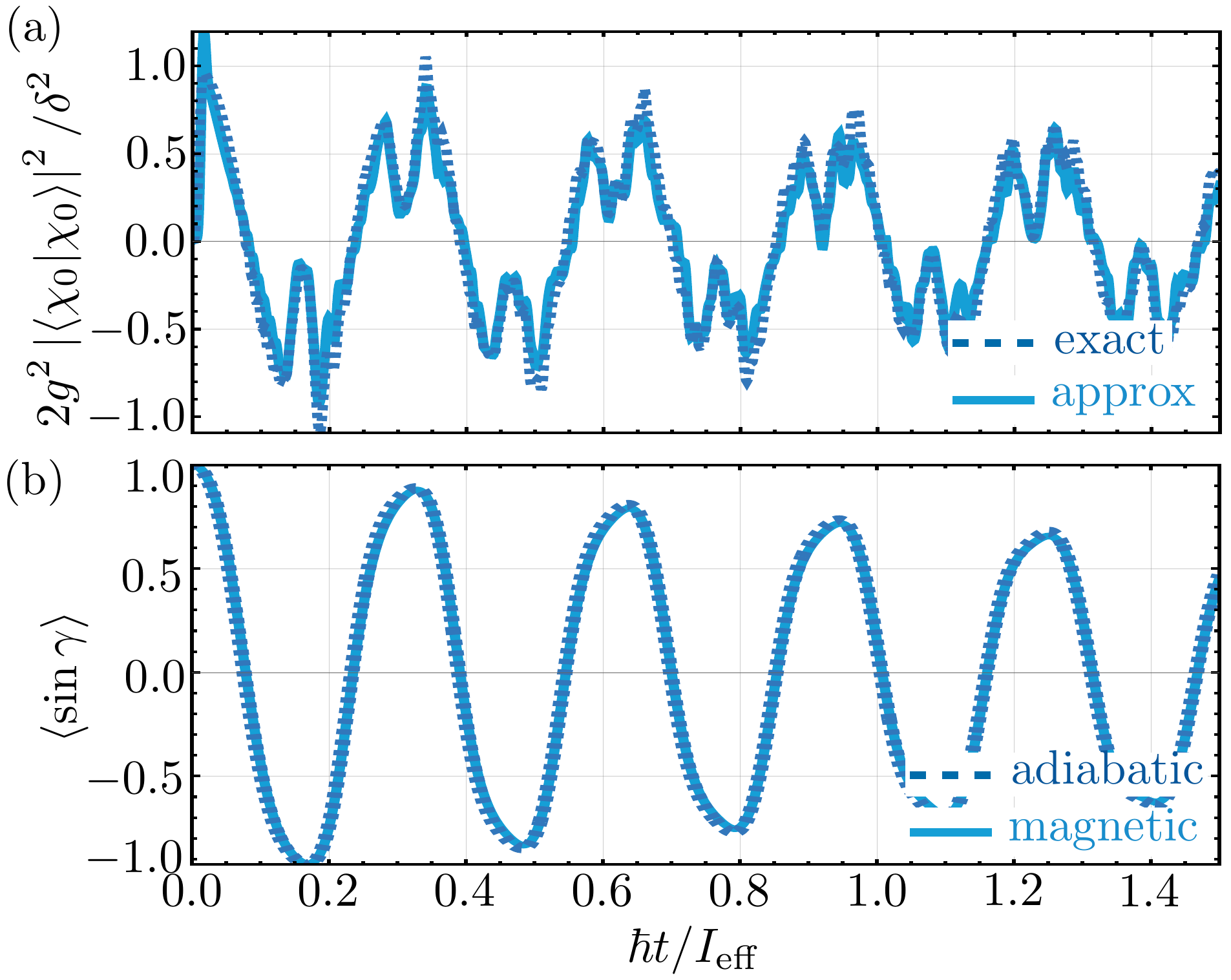}
\caption{(a) Occupation probability of the $|S_1=0\rangle$ spin state calculated from the exact dynamics given by Eq.~\eqref{eq:adiabaticham} (dark blue, dashed) compared to the approximated solution of Eq.~\eqref{eq:app_chi0} (light blue, solid). (b) Comparison of the $\gamma$-dynamics obtained from the adiabatic Hamiltonian~\eqref{eq:adiabaticham} (dark blue, dashed) and the magnetic spin Hamiltonian~\eqref{eq:magham} (light blue, solid). The nanodiamond has an ellipsoidal shape with semiaxes $l_3 = 200$\,nm and $l_1 = l_2 = 0.4 l_3$, $B=-0.5\,\mathrm{mT}$, and $\omega/2\pi=1\,\mathrm{MHz}$. The spin is initially in $\left|\sigma_x=-1\right>$ located at $\gamma=\pi/2$. We simulated the dynamics in the angular momentum basis with dimension $d_m = 1729$ and rescaled $\delta$ and $g$ by a factor $10^{-8}$ due to computational constraints.}
\label{fig:app_comparison2}
\end{figure}
To eliminate the spin zero state from the dynamics, we expand the adiabatic spin-rotor state in the eigenbasis of $S_1$ as
\begin{equation}\label{eq:app_chi}
    |\chi\rangle = \sum_{s = -1}^{+1} |\chi_s\rangle |S_1 = s\hbar\rangle,
\end{equation}
so that the Hamiltonian \eqref{eq:adiabaticham} yields the coupled equations
\begin{subequations}
 \begin{align}
    i\hbar\partial_t\left|\chi_{0}\right>=&\left[\dfrac{p_\gamma^2}{2I_{\rm eff}}-\hbar D_\mathrm{nv}\right]\left|\chi_{0}\right> \nonumber \\
    & -\dfrac{\hbar g}{\sqrt{2}}\sin\gamma\left(\left|\chi_{+1}\right>+\left|\chi_{-1}\right>\right),\\
  i\hbar\partial_t\left|\chi_{+1}\right>=&\left[\dfrac{p_\gamma^2}{2I_{\rm eff}}+\hbar g\cos\gamma\right]\left|\chi_{+1}\right>-\dfrac{\hbar g}{\sqrt{2}}\sin\gamma\left|\chi_{0}\right>,\\
  i\hbar\partial_t\left|\chi_{-1}\right>=&\left[\dfrac{p_\gamma^2}{2I_{\rm eff}}-\hbar g\cos\gamma\right]\left|\chi_{-1}\right>-\dfrac{\hbar g}{\sqrt{2}}\sin\gamma\left|\chi_{0}\right>.
 \end{align}
 \end{subequations}
For moderate $B$ fields, $|\omega - \gamma_0 B| \lesssim D_{\rm nv}$, the nonmagnetic state $|S_1 = 0\rangle$ remains far detuned from the magnetic states $|S_1 = \pm \hbar\rangle$ for all orientations $\gamma$. This implies that $i\hbar\partial_t  \left|\chi_{0}\right>\approx 0$ for all times, so that
\begin{equation}\label{eq:app_chi0}
 \left|\chi_{0}\right>\approx -\dfrac{\delta}{\sqrt{2} g} \sin\gamma\left( \left|\chi_{+1}\right> +\left|\chi_{-1}\right> \right),
\end{equation}
as seen in Fig.~\ref{fig:app_comparison2}(a). Inserting this approximation into the equations of motion for $|\chi_{\pm 1}\rangle$ allows us to identify the effective Hamiltonian in the magnetic spin subspace as \eqref{eq:magham}. We compare in Fig.~\ref{fig:app_comparison2}(b) the $\gamma$-dynamics around $\gamma=\pi/2$ of the adiabatic Hamiltonian with the derived magnetic Hamiltonian. Taylor expanding this Hamiltonian in linear order of $\eta = \gamma - \pi/2 \simeq 0$ gives
\begin{equation}\label{eq:maghamapp}
 H_\mathrm{mag} \simeq \frac{p_\gamma^2}{2I_{\rm eff}} + \frac{\hbar \delta}{2} \sigma_x - \hbar g  \sigma_z\eta = \frac{p_\gamma^2}{2I_{\rm eff}} + \hbar\Omega \mathrm{e}^{-i\phi \sigma_y} \sigma_x \mathrm{e}^{i\phi \sigma_y},
\end{equation}
with the angle
\begin{equation}
    \phi = \frac{1}{2} {\rm arctan} \left ( \frac{2 g\eta}{\delta} \right )\simeq \frac{g}{\delta}\eta
\end{equation}
 and the frequency
\begin{equation}
    \Omega= \frac{1}{2}\sqrt{\delta^2+4g^2\eta^2} \simeq \frac{\delta}{2} + \frac{g^2}{\delta} \eta^2.
\end{equation}
Here, we assume that $|g\eta| \ll |\delta|$. This condition remains fulfilled for all times for the harmonic oscillator ground state $|n_\eta = 0\rangle |\sigma_x = +1\rangle$ with harmonic frequency $\omega_\eta = (2\hbar g^2/I_\mathrm{eff}|\delta|)^{1/2}$ if
\begin{equation}
    \left |\frac{g}{\delta} \left (\frac{2 I_{\rm eff} \omega_\eta}{\hbar} \right )^{1/2} \right | = \left ( \frac{\hbar g^2}{8I_{\rm eff} |\delta|^3} \right )^{1/4} \ll 1.
\end{equation}
For $l_3 = 200\,\mathrm{nm}$, $l_1=l_2=0.3 l_3$, $\omega/2\pi=1\,\mathrm{MHz}$, and $B=-100\,\mathrm{mT}$, the above expression evaluates to $\sim 5\times 10^{-4}$, justifying the perturbative treatment.

\section{Spin-controlled rotational interference}\label{app:interference}

Starting from Eq.~\eqref{eq:adiabaticham} we derive the dispersive Hamiltonian given in Eq.~\eqref{eq:H_dispersive} by restricting the spin to the subspace $\{\left|\uparrow\right>,\left|\downarrow\right>\}=\{\left|S_1 = +\hbar\right>,\left|S_1 =-\hbar \right>\}$. Proceeding analogously as in  Sec.~\ref{app:spintransition}, we first harmonically approximate the system around the potential minimum $\gamma=0$ and subsequently adiabatically eliminate the $\left|\chi_{+1}\right>$ state of Eq.~\eqref{eq:app_chi}. By considering magnetic field strengths $g\simeq D_\mathrm{nv}$, it holds   $i\hbar\partial_t  \left|\chi_{+1}\right>\approx 0$ for all times, such that 
\begin{equation}
    \left|\chi_{+1}\right>\approx \dfrac{g}{\sqrt{2}(D_\mathrm{nv}+g)}\left|\chi_{0}\right>.
\end{equation}
The effective Hamiltonian for $\left|\chi_{\downarrow}=\chi_{-1}\right>$ and $\left|\chi_{\uparrow}=\chi_{0}\right>$ then follows as 
\begin{equation}
    H_2=\dfrac{p^2_\gamma}{2 I_\mathrm{eff}}+\dfrac{\hbar}{4}(g-\tilde{\delta})\gamma^2+\dfrac{\hbar}{2}\left[\Delta+\dfrac{(g+\tilde{\delta}) }{2}\gamma^2\right]\sigma_z-\dfrac{\hbar g}{\sqrt{2}}\sigma_x\gamma,
     \label{eq:app_H2}
\end{equation}
where we shifted the system by $-\hbar\Delta/2$ and defined
\begin{equation}
\tilde{\delta} = \dfrac{g^2}{D_\mathrm{nv}+g}.
\end{equation} The spin-oscillator interaction can be diagonalized by applying the unitary transformation $QH_2Q^\dagger$ with
\begin{equation}
  Q=\mathrm{exp}\left(i\dfrac{\pi}{2}\dfrac{|\mathbf{m}|\mathbf{e}_z+\mathbf{m}}{||\mathbf{m}|\mathbf{e}_z+\mathbf{m}|}\cdot\boldsymbol{\sigma}\right), 
    \label{eq:app_S2}
 \end{equation}
 where 
 \begin{equation}
 \mathbf{m}=\left[\dfrac{\Delta}{2}+\dfrac{(g+\tilde{\delta})}{4}\gamma^2\right]\mathbf{e}_z-\dfrac{g}{\sqrt{2}}\gamma\mathbf{e}_x
 \label{eq:app_m2}
 \end{equation} 
 denotes the vector of the local direction of the spin. The transformed Hamiltonian reads
\begin{align}
H_\mathrm{d}\simeq&\dfrac{p^2_\gamma}{2I_\mathrm{eff}}+\dfrac{\hbar(g-\tilde{\delta})}{4}\gamma^2\nonumber\\
&+\dfrac{\hbar\Delta}{2} \sigma_z+\hbar\left(\dfrac{g+\tilde{\delta}}{4}+\dfrac{ g^2}{\Delta}\right)\sigma_z\gamma^2,
\label{eq:app_Hdisp}
\end{align}
where we approximated the spin dependent potential satisfying the conditions of the  dispersive limit, i.e. 
\begin{equation}
    \dfrac{2 g^2}{\Delta^2}\left<\gamma^2\right>\ll 1, \; \dfrac{\tilde{\delta}+g}{\Delta}\left<\gamma^2\right>\ll 1,
\end{equation}
and neglected non-adiabatic corrections related to spin-flip transitions. Considering $|\Delta|\ll |g|$ we can approximate $\tilde{\delta}\approx g/2$ for which Eq.~\eqref{eq:H_dispersive} follows. Fig.~\ref{fig:app_comparison1}(b) shows the comparison of the $\gamma$-dynamics obtained from this Hamiltonian with the one of Eq.~\eqref{eq:adiabaticham} close to $\gamma=0$. From this Hamiltonian
the evolution operators for the oscillator state $|\chi_{\uparrow(\downarrow)}\rangle$ are obtained as
\begin{subequations}
\begin{align}
U_\uparrow &=\exp\left[-\dfrac{it}{\hbar}\left(\dfrac{p^2_\gamma}{2 I_\mathrm{eff}}+\dfrac{I_\mathrm{eff}}{2}\omega_\gamma^2\gamma^2\right)\right],\\
U_\downarrow&=\exp\left[-\dfrac{it}{\hbar}\left(\dfrac{p^2_\gamma}{2 I_\mathrm{eff}}+\dfrac{I_\mathrm{eff}}{2}\omega_\gamma^2\gamma^2-\dfrac{\hbar g}{4}\left(1+\dfrac{4g}{\Delta}\right)\gamma^2\right)\right].
\end{align}
    \label{eq:app_time_evolution}
\end{subequations}

The probability of measuring spin $\left | \uparrow\!(\downarrow)\right >$ after the interferometer sequence described in the main text, can be calculated as 
\begin{equation}
    P_{\uparrow\downarrow}=\dfrac{1}{2}\pm \dfrac{1}{4}\mathrm{Tr}\left[U_\downarrow^\dagger U_\uparrow^\dagger U_\downarrow U_\uparrow \rho_\mathrm{th}+U_\uparrow^\dagger U_\downarrow^\dagger U_\uparrow U_\downarrow \rho_\mathrm{th}\right].
\end{equation}
In the limit $T\ll\hbar \omega_\gamma/k_{\rm B}$, meaning that we only consider the ground state $|n_0\rangle$, the trace over the evolved thermal state can be calculated as 
\begin{align}
   I_\gamma&=2\mathrm{Re}\left[\langle n_0| U_\downarrow^\dagger U_\uparrow^\dagger U_\downarrow U_\uparrow|n_0\rangle\right] \nonumber \\
   &=2\mathrm{Re}\left[\left(1+\dfrac{(\Delta +4 g)^2(1-\mathrm{e}^{2i\tau\omega_\gamma})}{8 g (\Delta +2 g)}\sinh^2\zeta\right)^{-\frac{1}{2}}\right],
   \label{eq:app_probability}
\end{align} 
where we used $\Delta\ll g$ and introduced $\zeta=\sqrt{\hbar} g \tau/\sqrt{I_{\rm eff} \Delta}$.
For the second equality the Baker-Campbell-Hausdorff formula of SU(1,1) \cite{ban1992} was used to rewrite the evolution operators~\eqref{eq:app_time_evolution} \cite{rusconi2022spin}. For finite temperatures the expression for $I_\gamma$ becomes more complicated and can be found in the Supplement of Ref.~\cite{rusconi2022spin}.

\section{Rotation-induced microwave-drive detuning}\label{app:doppler}
Transforming the microwave drive from the lab frame into the rotor’s body frame introduces the rotational Doppler shift $\omega$ of the spin transition. In the rotating frame, the Barnett term shifts the spin splitting by $-\omega$ (in our sign convention), which precisely cancels the Doppler shift of the drive, leaving the microwave $\pi/2$ pulses resonant as the particle spins~\cite{chudo2015rotational}.

If the Paul trap is slightly asymmetric, $A_1=-1/2-\epsilon$, $A_2=-1/2+\epsilon$ with $\epsilon\ll 1$, the quadrupole potential acquires a weak $\alpha$-dependence. Linearizing around $\beta=\pi/2$ ($\xi=\pi/2-\beta$), valid under gyroscopic stabilization, yields the effective potential
\begin{align}
    V(\alpha,\beta)\approx &\dfrac{U_\mathrm{ac}^2(Q_1-Q3)^2}{36 \omega_\mathrm{ac}^2 d_0^4 I}\times\nonumber\\
    &\times\left[\dfrac{9}{4}\xi^2+3\epsilon\xi\cos(2\alpha)+\epsilon^2\sin^2(2\alpha)\right].
\end{align}
The resulting first-order change of  $p_\alpha$ is
\begin{equation}
    \dot{p}_\alpha^{(1)}=- \dfrac{U_\mathrm{ac}^2(Q_1-Q3)^2}{18 \omega_\mathrm{ac}^2 d_0^4 I}\left[\epsilon^2\sin(4\alpha)-3\epsilon\xi\sin(2\alpha)\right].
\end{equation}
Taking $\alpha\simeq\omega t$ at zeroth order and integrating over an integer number of full rotations, 
this correction averages to zero. Thus, the mean rotation frequency $\omega=p_\alpha/I$ remains unchanged to first order. The residual instantaneous modulation of the rotation frequency $\Delta \omega(t)$ scales as $\mathcal{O}(\epsilon^2)$ and $\mathcal{O}(\epsilon\xi)$, and is negligible for small asymmetries and small gyroscopic fluctuations ($\xi\ll1$) on timescales of the protocol.

\section{Decoherence channels }\label{app:decoherence}
In this section we quantify the dominant decoherence and heating mechanisms mentioned in the main text in more detail. All estimates are evaluated in the parameter regime of interest (rotation frequency $\omega/2\pi= 1\,\mathrm{MHz}$ and interferometer duration $\tau\leq 10–30\,\mu\mathrm{s}$). We show that each decoherence channel is negligible on these timescales.

\subsection{Magnetic-field noise}
Since the $\gamma$-rotation is not directly driven by the RF quadrupole, the dominant decoherence channel arises from magnetic dephasing of the NV center. Experiments on isolated  NV centers in nanodiamonds report coherence times of otder $T_2\sim 10\,\mu\mathrm{s}$ at room temperature~\cite{barry2020,chen2023extending, du2024single}. Thus $\gamma$-decoherence is primarily induced by fluctuations of the homogeneous magnetic field. 
 
The relevant noise source is Johnson-Nyquist current noise in the trap electrodes, which induces magnetic-field fluctuations that can be related to the electrode geometry via the Biot-Savart law. We model  fluctuations along $\mathbf{e}_z$ as $\delta \mathbf{B}(t)=\delta B(t)\mathbf{e}_z$ with Gaussian white noise $\left<\delta B(t)\delta B(t')\right>= A_\mathrm{fl}^2\delta(t-t')$, where $A_\mathrm{fl}$ denotes the magnetic-field noise amplitude. To quantify the residual effect on $\gamma$, we describe the NV as an orientation-dependent magnetic dipole $\mathbf{m}(\Omega)=-\gamma_0 \mathbf{S}$. For $\beta=\pi/2$, the fluctuations enter through the stochastic Hamiltonian 
\begin{equation}
    H_\mathrm{fl}(t)=-\mathbf{m}(\Omega)\cdot\delta\mathbf{B}(t)=\hbar\gamma_0\delta B(t)\cos\gamma.
\end{equation}
Expanding the interaction-picture propagator to second order in $H_\mathrm{fl}(t)$ and taking the ensemble average, yields the Markov master equation
\begin{equation}
\mathcal{L}\rho=\Gamma_B \left(\cos\gamma\rho \cos\gamma-\dfrac{1}{2}\{\cos^2\gamma,\rho\}\right),
\end{equation}
with $\Gamma_B=(\gamma_0  A_\mathrm{fl})^2$. To estimate the induced orientational decoherence, we evaluate $\left<\gamma\right|\mathcal{L}\rho\left|\gamma'\right>=(\gamma_0  A_\mathrm{fl})^2(\cos\gamma-\cos\gamma')^2/2$. For magnetic-field noise amplitudes of order $1\,\mathrm{nT}/\sqrt{Hz}$~\cite{mitchell2020colloquium}, and for angular separations of a few zero-point amplitudes $\gamma_\mathrm{zp}\sim(0.5-1)\times10^{-3}$, the resulting rate is $\lesssim 1\,\mathrm{Hz}$, well below the NV's intrinsic dephasing rate.

\subsection{Gas-collisions induced rotational decoherence}
Residual gas collisions induce rotational damping~\cite{monteiro2018optical,seberson2019parametric,kim2025magnetically}. The collision rate with gas particles of mass $m_\mathrm{g}$, temperature $T$ and pressure $P_\mathrm{g}$,  can be approximated for spheroidal particles as~\cite{martinetz2018gas,martinetz2023quantum}  
    \begin{equation}
        \Gamma_\mathrm{coll}=l_1(l_1+l_3)P_\mathrm{g}\sqrt{\dfrac{2\pi}{m_\mathrm{g}k_\mathrm{B}T}}.
    \end{equation}
For nitrogen ($m_\mathrm{g}=28\,\mathrm{u}$) at room temperature and $P_\mathrm{g}=10^{-8}\,\mathrm{mbar}$, a nanodiamond with $l_3=200\,\mathrm{nm}$, $l_1=l_2=0.3 l_3$ experiences collisions at rate $ \Gamma_\mathrm{coll}\approx 2.8\times 10^{3}/\mathrm{s}$, corresponding to $\approx 0.03$ collisions during a $10\,\mu\mathrm{s}$ protocol. Collisional decoherence is therefore negligible and can be fully suppressed under ultra-high vacuum ($P_\mathrm{g}\leq10^{-8}\,\mathrm{mbar}$).

\subsection{Dipole noise and electric-field noise}
Highly charged spheroidal particles tend to have negligible electric dipole moments~\cite{martinetz2020quantum} which are experimentally unobserved~\cite{delord2020spin,perdriat2024rotational}. Thus, the dominant electric noise source is quadrupole-field noise in the RF trap, known to cause motional heating~\cite{martinetz2022surface,glikin2025probing}. In our geometry, the $\gamma$-mode does not couple to the quadrupole field to leading order because the RF field primarily excites the $\beta$-mode (which is eliminated in the effective model). Consequently, quadrupole-field noise perturbs $\beta$, while leaving $\gamma$ unaffected. 

Further, in a Paul trap, the RF quadrupole potential is arranged such that the homogeneous electric field vanishes at the trap center, eliminating any leading order Stark coupling to the NV spin. Small static offsets of the particle from the center, due to stray DC fields, gravity, or residual micromotion, generate a local AC field proportional to the displacement. For the sub-micron offsets typical of compensated traps, these fields are strongly reduced by diamond’s dielectric screening ($\varepsilon_\mathrm{r}\approx5.7$) and remain far below the NV electric-field sensitivity ($\sim17\,$Hz cm/V)~\cite{dolde2011electric,doherty2012theory}. Residual Stark shifts are therefore negligible compared to both the NV’s intrinsic decoherence rate and the spin-rotation coupling scale.

\subsection{Black-body radiation}
Thermal photons cause orientational decoherence through absorption and emission processes. In thermal equilibrium, the radiation field couples to the particle via the complex polarizability tensor $\upalpha(\omega_\mathrm{ph})=\varepsilon_0 V\upchi(\omega_\mathrm{ph})$, with $V$ the particle volume, $\varepsilon_0$ the vacuum permittivity, and  $\upchi(\omega_\mathrm{ph})$ the susceptibility tensor. Its imaginary part $\upalpha''(\omega_\mathrm{ph})$ accounts for electromagnetic absorption. The decay of orientational coherences between two configurations $\gamma$ and $\gamma'$ is governed by the localization rate $ F({\gamma,\gamma'})$~\cite{schaefer2024decoherence},
    \begin{equation}   F({\gamma,\gamma'})=\int_0^\infty\dfrac{\mathrm{d}\omega_\mathrm{ph}\omega_\mathrm{ph}^3}{3\pi^2 c^3 \varepsilon_0}\bar{n}(\omega_\mathrm{ph})\mathrm{Tr}\left[\upalpha''(\omega_\mathrm{ph})\left(\mathbb{1}-\mathrm{R}^T\mathrm{R}'\right)\right],
    \end{equation}
with $\mathrm{R}^T\mathrm{R}'\equiv\mathrm{R}^T(\gamma)\mathrm{R}(\gamma')$, $c$ the speed of light, and $\bar{n}(\omega_\mathrm{ph})=[\mathrm{exp}(\hbar\omega/k_\mathrm{B}T)-1]^{-1}$ the thermal photon occupation number. Assuming the polarizability to be frequency-independent, $\upalpha''(\omega_\mathrm{ph})=\alpha_\perp''\mathbb{1}+(\alpha_\parallel''-\alpha_\perp'')\mathbf{e}_z\otimes \mathbf{e}_z$, where $\alpha_\perp''$, $\alpha_\parallel''$  are the principal values along the minor and major axes, respectively, the localization rate reduces to
 \begin{equation}
     F({\gamma,\gamma'})=\Gamma_\mathrm{ph}[1-\cos(\gamma-\gamma')],
 \end{equation}
with the photon emission rate
 \begin{equation}
     \Gamma_\mathrm{ph}=\dfrac{2 \pi^2 \alpha_\perp'' k_\mathrm{B}^4 T^4 }{45c^3 \hbar^4 \varepsilon_0}.
 \end{equation}
 Under ambient conditions ($T\simeq 300$ K), the emission rate is $\Gamma_\mathrm{ph}/2\pi\approx 7.2\,\mathrm{MHz}$ for a transverse polarizability of $\alpha''_\perp \simeq 1\times 10^{-32}\,\mathrm{C}\mathrm{m}^2/\mathrm{V}$~\cite{schaefer2024decoherence}. The relevant decoherence depends on the angular separation in the superposition, as determined by the amount of squeezing.  The interferometer produces superpositions of a few zero-point amplitudes $\gamma_\mathrm{zp}\sim(0.5-1)\times10^{-3}$, so that $F(\gamma,\gamma')/2\pi\lesssim 30\,\mathrm{Hz}$, well below the interferometer timescale.

Rayleigh scattering of thermal photons does not contribute because the  $\gamma$-rotation is symmetric  under the angular scattering distribution~\cite{papendell2017quantum}.

%\bibliographystyle{myapsrev}
%\bibliography{refs} 

%

\end{document}